\documentclass[fleqn,usenatbib]{mnras}

\usepackage{newtxtext,newtxmath}
\usepackage{hyperref}
\usepackage{booktabs}

\usepackage[T1]{fontenc}

\DeclareRobustCommand{\VAN}[3]{#2}
\let\VANthebibliography\thebibliography
\def\thebibliography{\DeclareRobustCommand{\VAN}[3]{##3}\VANthebibliography}

\usepackage{graphicx}	
\usepackage{amsmath}	
\usepackage{orcidlink}
\usepackage{lipsum}
\usepackage{widetext}
\newcommand*{\mailto}[1]{\href{mailto:#1}{#1}}

\title[Non-universal PPSD in Symphony]{The Non-universal Pseudo Phase-Space Density Profiles of Symphony Host haloes}

\author[Feng \& Nadler et al.]{
Bocheng Feng$^{\orcidlink{0009-0008-2453-2497}{}1}$\thanks{\mailto{bochengfeng.phy@stu.pku.edu.cn}},Ethan O.~Nadler$^{\orcidlink{0000-0002-1182-3825}2}$\thanks{\mailto{enadler@ucsd.edu}},
S.~Peng Oh$^{\orcidlink{0000-0002-1013-4657}3}$,
and Suoqing Ji$^{\orcidlink{0000-0001-9658-0588}4,5}$
\\
$^{1}$Department of Astronomy, School of Physics, Peking University, Beĳing 100871, China\\
$^{2}$Department of Astronomy \& Astrophysics, University of California, San Diego, La Jolla, CA 92093, USA\\
$^{3}$Department of Physics, University of California, Santa Barbara, Santa Barbara, CA 93106, USA\\
$^{4}$Center for Astronomy and Astrophysics and Department of Physics, Fudan University, Shanghai 200438, China\\
$^{5}$Key Laboratory of Nuclear Physics and Ion-Beam Application (MOE), Fudan University, Shanghai 200433, P.R.China\\
}

\pubyear{2025}

\begin{document}
\label{firstpage}
\pagerange{\pageref{firstpage}--\pageref{lastpage}}
\maketitle

\begin{abstract} 
Cosmological N-body simulations have long suggested that the pseudo phase-space density (PPSD), $\rho/\sigma^3$, of cold dark matter haloes follows the universal relation $\rho/\sigma^3 \propto r^{\chi}$, with $\chi \approx -1.875$, as predicted by spherical secondary-infall similarity solutions. This power law appears to hold despite the fact that neither the density $\rho(r)$ nor velocity dispersion $\sigma(r)$ follow universal power law relations individually, even at fixed mass. We analyze 246 host haloes from the \textit{Symphony} suite of high-resolution cosmological zoom-in simulations, to consistently measure PPSD profiles across host masses from $10^{11}$ to $10^{15} M_\odot$. We find that the PPSD systematically deviates from a power law, and that haloes with larger deviations from Jeans equilibrium systematically develop steeper average PPSD slopes. This result suggests that the PPSD is not universal; instead, it is linked to a halo's degree of dynamical equilibrium, which is ultimately set by halo formation history. As a result, we show that secondary halo properties such as concentration and accretion rate inherit significant correlations with the PPSD slope. Moreover, our hosts' PPSD profiles are remarkably consistent with predictions from one-dimensional self-similar fluid collapse models, indicating that three-dimensional structure, velocity anisotropy, and filamentary accretion all play negligible roles in shaping the PPSD. Thus, we argue that the PPSD is shaped by mass assembly alone, and that its non-universality reflects the diversity of halo growth histories.
\end{abstract}




\section{INTRODUCTION}
\label{intro}

The inner structure of dark matter haloes is a key prediction of the cold dark matter (CDM) paradigm. In the standard cosmological picture, haloes form through hierarchical gravitational collapse, continuously growing via mergers and the smooth accretion of surrounding material. Despite the complexity of this process, cosmological N-body simulations reveal that haloes share remarkably similar internal structures \citep{1974ApJ...187..425P, 1978MNRAS.183..341W,Huss_1999, van_den_Bosch_2002}. Cosmological \textit{N}-body simulations show that the spherically averaged density profiles of haloes are well described by simple functional forms such as the Navarro--Frenk--White (NFW; \citealt{1996ApJ...462..563N, Navarro_1997}) or Einasto profiles \citep{1965TrAlm...5...87E}, exhibiting only mild variations across halo mass and redshift
\citep{2010arXiv1010.2539D, Hjorth_2015, Wang_2020}. This structural similarity has motivated extensive efforts to understand the physics governing the emergence of universal halo structure (e.g., see \citealt{Zavala_2019} for a review).

Beyond their spatial structure, dark matter haloes also display unexpected universality in phase space. A particularly striking aspect of halo dynamics is the so-called pseudo phase-space density profile (PPSD; \citealt{2001ApJ...563..483T}), defined as
\begin{equation}
    Q(r) \equiv \rho(r)/\sigma^3(r),
\end{equation} 
where $\rho$ is the dark matter mass density and $\sigma$ is the total or radial velocity dispersion (we will discuss this choice below). The PPSD connects spatial and kinematic structure: in collisionless systems, the fundamental invariant is the fine-grained phase-space density $f(\mathbf{x},\mathbf{v})$, which is conserved according to Liouville’s theorem~\citep{2008gady.book.....B}. Although $f$ cannot be directly observed or measured in simulations, the PPSD serves as a coarse-grained proxy of the true phase-space density $f$ and has been shown to trace the dominant structures in phase space \citep{Maciejewski_2009}.

Early $N$-body simulations revealed that the PPSD is surprisingly universal: for simulated CDM haloes, $Q(r)$ follows a power law $Q(r) \propto r^{\chi}$, with $\chi \approx -1.875$, over 2 decades in radius and with very little variation across haloes \citep{2001ApJ...563..483T,2005MNRAS.363.1057D, 2010MNRAS.402...21N, Ludlow_2010, 2011MNRAS.415.3895L, Colombi_2021}. This behavior is particularly striking because neither $\rho(r)$ nor $\sigma(r)$ follow single power laws themselves, and both vary from halo to halo, even at fixed mass. The apparent universality of $\rho/\sigma^3$ therefore raises a fundamental question:  
\emph{what dynamical mechanism enforces this power-law behavior in collisionless, hierarchically growing systems?}  
For example, does the PPSD point to an additional integral of motion for self-gravitating systems, or is it simply an emergent coincidence?

The puzzle deepens when considering that the measured slope $\chi \approx -1.875$ matches the value predicted by self-similar secondary infall \citep{1985ApJS...58...39B, williams2005phasespacedensitydistributiondark}, an idealized model describing the collapse of matter onto a point mass in an Einstein–de Sitter universe \citep{1985ApJS...58...39B, williams2005phasespacedensitydistributiondark}. Real haloes, however, assemble through mergers \citep{2008MNRAS.383..546C}, develop triaxial shapes \citep{2006MNRAS.367.1781A}, and exhibit anisotropic velocity distributions \citep{2006NewA...11..333H}, none of which resemble the assumptions of the self-similar model. Why such a simplified analytic model reproduces the PPSD scaling observed in full cosmological simulations remains one of the most intriguing open questions in halo dynamics.

Beyond its theoretical importance, the PPSD has also been invoked as an observational diagnostic of the innermost density profiles of CDM haloes. If $Q(r)$ indeed follows a nearly universal power law, then empirical determinations of $\sigma(r)$---from stellar dynamics in galaxies, globular cluster kinematics, or distortions in strong-lensing arcs---can be used to infer the underlying density profile by extrapolating\footnote{However, such extrapolations must be treated with caution in halo centers, where baryons may reshape the potential and flatten the PPSD profiles via adiabatic contraction or feedback~\citep{1986ApJ...301...27B,Gnedin_2004, Faltenbacher_2007, 2010MNRAS.405.2161D,2012MNRAS.421.3464P, Cintio_2014}.}. On group and cluster scales, joint analyses that combine dark matter kinematics with the thermodynamic properties of the baryonic gas have probed the PPSD \citep{Pratt_2006, Voit_2005, 2009ApJS..182...12C, Capasso_2018}. For instance, \citet{Faltenbacher_2007} demonstrated that the radial PPSD and entropy profiles inferred for galaxy clusters are consistent with power laws and exhibit comparable slopes. More recently, \citet{Biviano_2023} analyzed the kinematics of 54 clusters from the Wide-field Nearby Galaxy-cluster Survey (WINGS) and found that their inferred PPSD profiles are consistent with an approximate power-law, providing observational support for the numerical prediction.

Despite these motivations, a growing body of work also challenges the strict universality of the PPSD \citep{williams2005phasespacedensitydistributiondark, Ludlow_2010, 2011MNRAS.415.3895L, Popolo_2011, Nadler_2017, Arora_2020}. Using one-dimensional fluid simulations, \citet{Nadler_2017} argued that deviations from equilibrium, which could arise from recent accretion or incomplete virialization, naturally produce departures from the $\chi \approx -1.875$ power law. Meanwhile, \citet{Arora_2020} showed that power-law PPSD solutions are generally incompatible with the structure of the Jeans equation for dynamically relaxed systems. However, these lines of argument are idealized; they both invoke spherical symmetry, and additionally assume either the fluid approximation or dynamical equilibrium. In contrast, cosmological haloes grow within a rich, hierarchical phase-space environment \citep{Ludlow_2010}, and it remains unclear whether these idealized expectations for the PPSD carry over to realistic structure formation.

These issues highlight a fundamental gap in our understanding: \textit{despite decades of work, we still lack a cosmologically grounded explanation for why PPSD profiles follow a near-universal scaling—and under what conditions this behavior breaks down.} This motivates a systematic reanalysis of PPSD structure using cosmological simulations that capture hierarchical growth and mergers \citep{2012MNRAS.419.1576P,Ludlow_2014}, as well as internal dynamics such as velocity anisotropy \citep{2006NewA...11..333H} and triaxiality \citep{2002ApJ...574..538J}. In this work, we therefore aim to quantify the degree of universality in the PPSD—both within individual haloes and across the halo population—and to identify the physical processes that set its shape. To achieve this, we use the \textit{Symphony} suite of high-resolution CDM zoom-in simulations spanning host halo masses from $10^{11}$ to $10^{15},M_\odot$ \citep{Nadler_2023}. Because of its large dynamic range and the fact that each halo is resolved with a similar number of particles within its virial radius (despite spanning a wide range of halo masses), \textit{Symphony} provides an ideal dataset for investigating how $Q(r)$ depends on halo mass, assembly history, dynamical state, and secondary halo properties. By combining detailed measurements of density, velocity dispersion, and anisotropy profiles, we quantify deviations from Jeans equilibrium and show how these shape PPSD profiles. As a dark matter-only suite, \textit{Symphony} lacks baryonic processes such as adiabatic contraction and feedback-driven core formation. However, this collisionless framework allows us to robustly test the predictions of secondary infall and self-similar collapse models in realistic cosmological environments while isolating the effects of gravitational assembly from complex baryonic physics.

This paper is organized as follows. In Section~\ref{sec:simulations}, we describe our zoom-in simulations and analysis techniques. In Section~\ref{sec:ppsd}, we measure the density, velocity dispersion, velocity anisotropy, and PPSD profiles across \textit{Symphony} suites. In Section~\ref{sec:non_universality}, we study how deviations from Jeans equilibrium shape PPSD profiles, the redshift evolution of the PPSD, and correlations between secondary halo properties and PPSD slopes. In Section~\ref{sec:fluid_collapse}, we compare our PPSD results with the \citet{Nadler_2017} one-dimensional fluid simulations. In Section~\ref{sec:summary}, we summarize our main results and discuss.

\section{Symphony Simulations and Analysis Methods}
\label{sec:simulations}

\begin{figure*}
    \centering
    \includegraphics[width=1\linewidth]{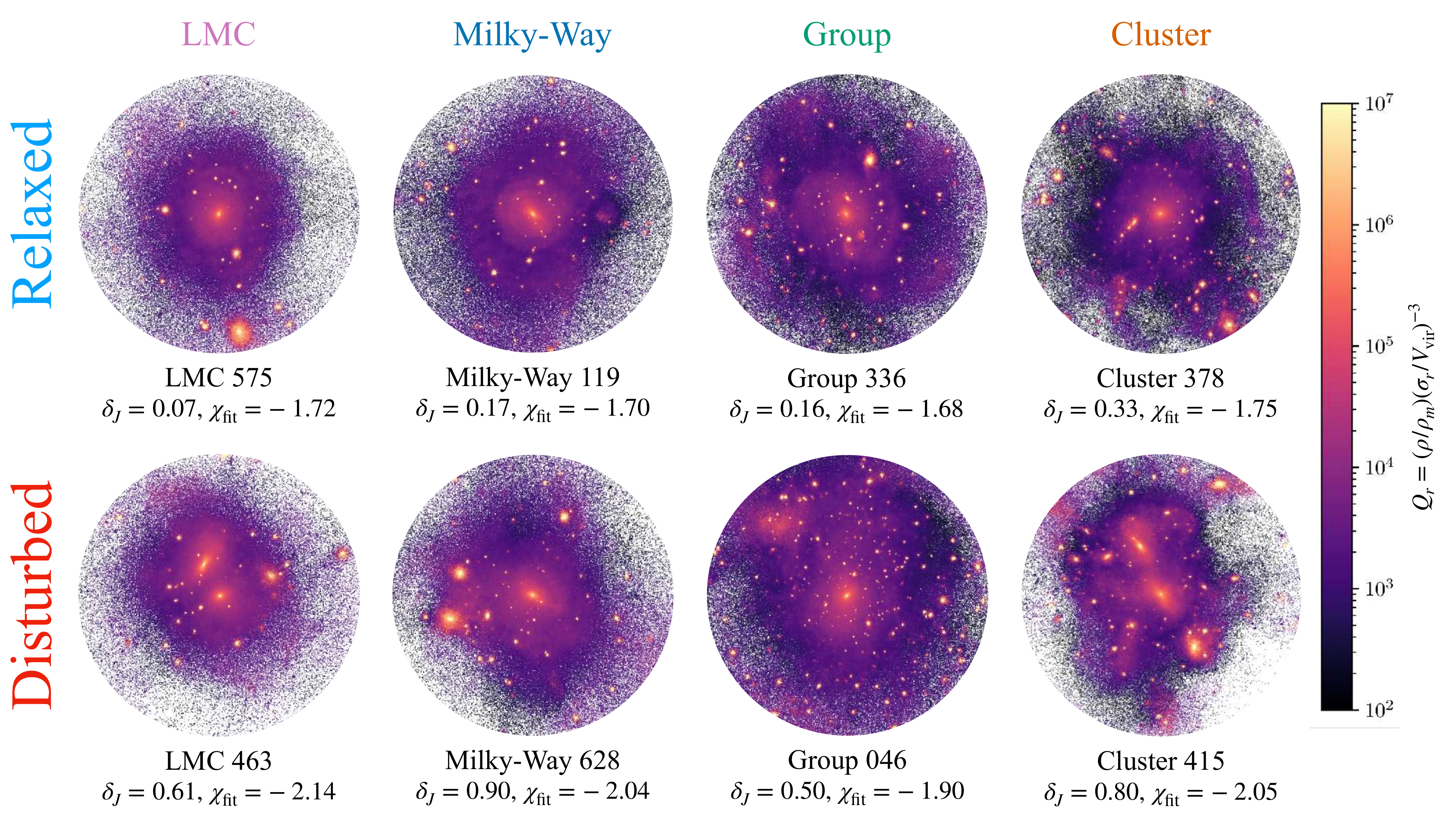}
    \caption{Projected radial pseudo phase-space density (PPSD) maps of representative host haloes at $z=0$ from the LMC, Milky-Way, Group, and Cluster suites in our simulations, shown across columns. The top row shows relaxed systems close to Jeans equilibrium, while the bottom row shows disturbed systems far from Jeans equilibrium in the corresponding suites. The departure from equilibrium is quantified by the Jeans deviation parameter $\delta_J$ (see Section~\ref{sec:jeans_dev} for details). For each host, the best-fit average radial PPSD slope, $\chi_{\rm fit}$, is also indicated in the corresponding panel. Each visualization is a slice of thickness $0.2R_{\rm vir}$ extending out to the virial radius. 
}
    \label{fig:ppsd_visual}
\end{figure*}

We use 246 cosmological, CDM host haloes in \textit{Symphony} simulations, a compilation of five dark matter--only zoom-in suites: 
\begin{itemize}
    \item LMC-mass hosts ($10^{11} M_{\odot}$; the ``LMC'' suite);
    \item Milky Way-mass hosts ($10^{12} M_{\odot}$; the ``Milky Way'' suite);
    \item Group-mass hosts ($10^{13} M_{\odot}$; the ``Group'' suite);
    \item Low-mass cluster hosts ($4\times 10^{14} M_{\odot}$; the ``L-Cluster'' suite);
    \item Cluster-mass hosts ($10^{15} M_{\odot}$; the ``Cluster'' suite)
\end{itemize}
Across the entire mass range, each host halo is resolved with more than $10^6$ particles within the virial radius at $z=0$, and the highest-resolution regions are resolved with a Plummer-equivalent gravitational softening length of $\approx 9 \times 10^{-4}R_{\rm vir}$, on average, where $R_{\rm vir}$ is the host halo virial radius. Below, we provide a brief summary of simulation parameters and our analysis techniques. More details about the \textit{Symphony} simulations can be found in \citet{Nadler_2023}.

\subsection{Cosmological parameters}

The \textit{Symphony} compilation adopts slightly different cosmological parameters across its five suites, reflecting the specifications of their parent simulations. The LMC, Milky-Way, and Group suites are drawn from a $1024^3$-particle,
$125 h^{-1}~\mathrm{Mpc}$-box run in a flat $\Lambda$CDM cosmology with $h=0.7$, $\Omega_{\rm m}=0.286$, $\Omega_{\Lambda}=0.714$, $\sigma_{8}=0.82$, and $n_{s}=0.96$~\citep{Hinshaw_2013,Mao_2015}. These cosmological parameters correspond to a virial overdensity of $\Delta_{\rm vir}\simeq 99$ at $z=0$ \citep{1998ApJ...495...80B}. 

The L-Cluster suite originates from a $1024^3$-particle, $1 h^{-1}~\mathrm{Gpc}$-box parent run with cosmological parameters $h=0.7$, $\Omega_{\rm m}=0.3$, $\Omega_{\Lambda}=0.7$, $\sigma_{8}=0.85$, and $n_{s}=0.96$, corresponding to $\Delta_{\rm vir}\simeq 101$~\citep{Bhattacharyya_2022}.  

Finally, the Cluster suite is based on the Carmen simulation \citep{McBride_2009} of side length $1 h^{-1}~\mathrm{Gpc}$ with $1024^3$ particles per side, which adopts cosmological parameters $h=0.7$, $\Omega_{\rm m}=0.25$, $\Omega_{\Lambda}=0.75$, $\sigma_{8}=0.8$, and $n_{s}=1.0$, corresponding to $\Delta_{\rm vir}\simeq 94$ at $z=0$~\citep{Wu_2013,Wu_2013_2}. These small cosmological differences among suites do not impact our results but should be kept in mind for cross-suite comparisons.

Across the entire \textit{Symphony} compilation, the particle mass scales approximately as $m_{\rm part} \simeq 3\times 10^{-7} M_{\rm host}$, allowing well-resolved measurements of host density and kinematic profiles over four decades in host halo mass. The comoving Plummer-equivalent gravitational softening length is set to $\epsilon \simeq 9\times 10^{-4} R_{\rm vir}$, on average, which avoids both artificial suppression of central densities from excessive softening and integration errors from insufficient softening~\citep{Nadler_2023}. We list the main properties of each suite in Table~\ref{tab:Symphony}.

\begin{table*}
  \centering
  \caption{Numerical properties of the five \textit{Symphony} suites. The first column lists the name of suite, the second column lists the number of zoom-in simulations we use, the third column lists the the median and standard deviation of the host halo virial mass, the fourth column lists the dark matter particle mass in the highest-resolution zoom-in region, the fifth column lists the comoving Plummer-equivalent force softening length, and the sixth column lists the convergence radius $2.8\epsilon$. Note that we omit 16 simulations from the Cluster suite, relative to the 96 presented in \citet{Nadler_2023}, due to corrupted particle data.}
  \label{tab:Symphony}
  \begin{tabular}{lcccccc}
    \toprule
    Zoom-in Suite & $N_{\rm sim}$ & $M_{\rm host}$ ($M_\odot$) & $m_{\rm p}$ ($M_\odot$) & $\epsilon$ (pc $h^{-1}$) & $r_{\mathrm{conv}}$ (pc $h^{-1}$)\\
    \midrule
    LMC            & 39 & $10^{11.02\pm0.05}$ & $5.0\times 10^{4}$  & 80 & 224\\
    Milky Way      & 45 & $10^{12.09\pm0.02}$ & $4.0\times 10^{5}$  & 170 & 476\\
    Group               & 49 & $10^{13.12\pm0.11}$ & $3.3\times 10^{6}$  & 360 & 1008\\
    L-Cluster    & 33 & $10^{14.62\pm0.11}$ & $2.2\times 10^{8}$  & 1200 & 3360\\
    Cluster             & 80 & $10^{14.96\pm0.03}$ & $1.8\times 10^{8}$  & 3295 & 9226\\
    \bottomrule
  \end{tabular}
\end{table*}

\subsection{Analysis Techniques}

\subsubsection{Density, Velocity and PPSD Profiles}

To compute the profiles of the spherically averaged pseudo phase-space density (PPSD) as a function of radius, we divide each host into 40 logarithmically-spaced spherical shells spanning $10^{-3}R_{\text{vir}}$ to $1.5R_{\text{vir}}$, where $R_{\text{vir}}$ denotes the virial radius of each host. All particles within each shell are included in the calculation (i.e., including substructures). The average density profile $\rho(r)$ in each bin is obtained by dividing the total mass within the shell by its volume. The radial velocity dispersion profile, $\sigma_{\text{rad}}(r)$, is measured in the host rest frame as the average radial velocity dispersion of particles in each shell. Figure~\ref{fig:ppsd_visual} shows examples of PPSD profiles in different suites across columns; we further illustrate dynamically``relaxed'' versus ``disturbed'' haloes within each suite in different rows. As we will show below, the magnitude of the deviation from Jeans equilibrium correlates with properties of the PPSD profiles.

The total velocity dispersion is defined as the quadratic sum of the dispersions in the three Cartesian components,
$\sigma_{\text{tot}}^2 = \sigma_{x}^2 + \sigma_{y}^2 + \sigma_{z}^2$,
which is equivalent to twice the specific kinetic energy. The velocity anisotropy parameter is then defined as
\begin{equation}
    \beta(r) \equiv 1 - \frac{\sigma_{\text{tan}}^2(r)}{2\sigma_{\text{rad}}^2(r)}, \label{eq:vel_anisotropy}
\end{equation}
where the tangential velocity dispersion is given by $\sigma_{\text{tan}}^2 = \sigma_{\text{tot}}^2 - \sigma_{\text{rad}}^2$. With this definition, $\beta = 0$ corresponds to an isotropic velocity distribution, while $\beta = 1$ describes purely radial orbits.

\begin{figure*} 
    \centering
    \includegraphics[width=1\linewidth]{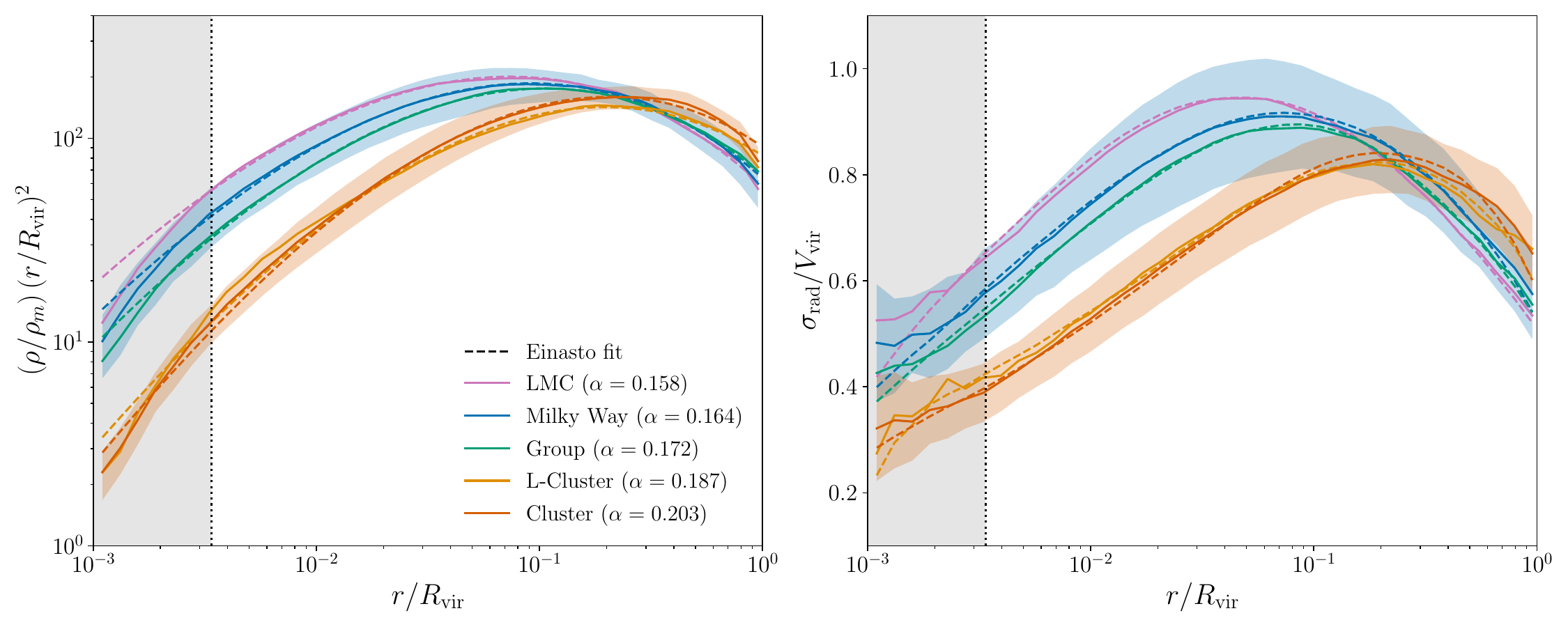}
    \caption{Density (left) and radial velocity dispersion (right) profiles of host haloes in \textit{Symphony} suites. Solid lines show suite averages, with radii scaled by $R_{\text{vir}}$, densities scaled by $\rho_m$ (the mean matter density at $z=0$), and velocity dispersions scaled by $v_{\text{vir}}$ (the host virial velocity). Densities are multiplied by $(r/R_{\text{vir}})^2$, to enhance the dynamic range. Dashed lines in the left panel show the best-fit Einasto density profiles, while dashed lines in the right panel are obtained by solving the Jeans equation for the corresponding Einasto models using the measured velocity anisotropy profiles. Shaded regions indicate the 68\% halo-to-halo scatter within the Milky Way and Cluster suites. The dotted vertical line marks the most conservative convergence radius, 2.8$\epsilon$, across all Symphony suites; throughout our analyses, we do not use measurements in the grey region $r/R_{\mathrm{vir}} < 2.8\epsilon$.
}
    \label{fig:density_velocity}
\end{figure*}

We consider both the total PPSD, defined as $Q_{\rm tot}(r) \equiv \rho / \sigma^3_{\text{tot}}$, and the radial PPSD, $Q_r(r) \equiv \rho / \sigma^3_{\text{rad}}$. In the following analysis we focus primarily on $Q_r(r)$, since the radial velocity dispersion directly enters the Jeans equation. We demonstrate in Appendix~\ref{sec:dimensional_ppsd} that the choice between $Q_{\rm tot}$ and $Q_r$ does not affect our results.

\subsubsection{PPSD slope smoothing}
\label{sec:ppsd_slope_smooth}

We compute smooth derivatives of the PPSD profiles using the \textsc{pynumdiff} package~\citep{ParamOptimizationDerivatives2020,PyNumDiff2022}. 
In particular, we adopt the \texttt{constant\_jerk} variant of the Kalman smoothing algorithm, which assumes a locally constant third derivative (jerk) of the signal. As demonstrated in Appendix~\ref{sec:smooth_derivative}, this method can reliably recover underlying power-law PPSD profiles within the radial range $6\times10^{-3}R_{\text{vir}}$–$1.1R_{\text{vir}}$. We therefore adopt this radial range when estimating best-fit average PPSD slopes, hereafter denoted as $\chi_{\rm fit}$.

To enable stacking of halo profiles across different mass scales, we normalize the radial coordinate by each host halo's virial radius, $r \rightarrow r/R_{\rm vir}$, the density by the mean matter density at $z=0$ in each suite, $\rho \rightarrow \rho/\rho_m$, and the velocity dispersion by each host halo's virial velocity, $\sigma \rightarrow \sigma/V_{\rm vir}$. Accordingly, the PPSD is expressed in units of $\rho_m / V_{\rm vir}^3$. In the following, all profiles and derived quantities are presented using this normalization unless stated otherwise. We also provide an empirical fit to the PPSD amplitude as a function of halo mass in Appendix~\ref{sec:dimensional_ppsd}.

\section{Symphony Host Halo Profiles}
\label{sec:ppsd}

In this section, we present host halo density, velocity dispersion, velocity anisotropy and PPSD profiles. We discuss our main results in Section~\ref{sec:non_universality}.

\subsection{Density and velocity dispersion profiles}

Figure~\ref{fig:density_velocity} shows the mean density and radial velocity dispersion profiles (measured in the host rest frame) of host haloes at $z=0$ in each simulation suite. The dotted line marks the ``convergence radius,'' $2.8 \epsilon$ \citep{Ludlow_2019}, shown here for the Cluster suite because it is the largest (and thus most conservative) convergence radius across all suites when expressed in units of $R_{\mathrm{vir}}$. Shaded regions denote the $\pm 1\sigma$ halo-to-halo scatter within each suite. For clarity, the density profiles are multiplied by $r^2$, which highlights the radius where the logarithmic slope, $\gamma(r) \equiv \mathrm{d}\ln\rho(r) / \mathrm{d}\ln r$, reaches $\gamma = -2$.

In the left panel of Figure~\ref{fig:density_velocity}, the dashed lines represent fits of the Einasto profile \citep{1965TrAlm...5...87E} to the mean density profiles of each suite. This three-parameter model is defined as
\begin{equation}
    \ln\!\left(\frac{\rho}{\rho_{-2}}\right)
    = -\frac{2}{\alpha} \left[\left(\frac{r}{r_{-2}}\right)^{\alpha} - 1\right], \label{eq:einasto}
\end{equation}
where $\rho_{-2} \equiv \rho(r_{-2})$ and $r_{-2}$ denote the density and radius at which the logarithmic slope equals $-2$, respectively. The shape parameter $\alpha$ controls the curvature of the density profile. As expected, the Einasto model provides a more accurate description of our host haloes than the Navarro--Frenk--White (NFW; \citealt{1996ApJ...462..563N, Navarro_1997}) profile. In particular, the additional degree of freedom introduced by $\alpha$ is essential to capture the curvature in the measured density profiles, consistent with previous results (e.g. \citealt{Navarro_2004, Merritt_2005, Merritt_2006, Gao_2008, 2008MNRAS.388....2H, Navarro_2009, 2011MNRAS.415.3895L}). Furthermore, as shown in the left panel of Figure~\ref{fig:density_velocity}, the best-fit values of $\alpha$ systematically increase with host halo mass, and this trend becomes steeper at higher masses. Note that the inferred values of $\alpha$ for each suite are consistent with those reported in earlier studies (e.g. \citealt{Ludlow_2013, Dutton_2014}).

To provide a practical method for fitting velocity dispersion profiles, we also construct $\sigma_{\mathrm{rad}}$ predictions based on the Einasto density profile. In the right panel of Figure~\ref{fig:density_velocity}, the dashed lines show the velocity dispersion profiles constructed from the corresponding Einasto models in the left panel. These constructed profiles are obtained by solving the Jeans equation using the measured velocity anisotropy profiles for each suite, assuming an Einasto density profile.\footnote{We assume that the velocity anisotropy goes to zero outside the virial radius when constructing these solutions.} Velocity dispersion profiles also exhibit systematic variations in shape across the \textit{Symphony} suites, and the Einasto-based solutions reproduce them accurately over the full range of host halo masses considered.

\begin{figure*}
    \centering
    \includegraphics[width=1\linewidth]{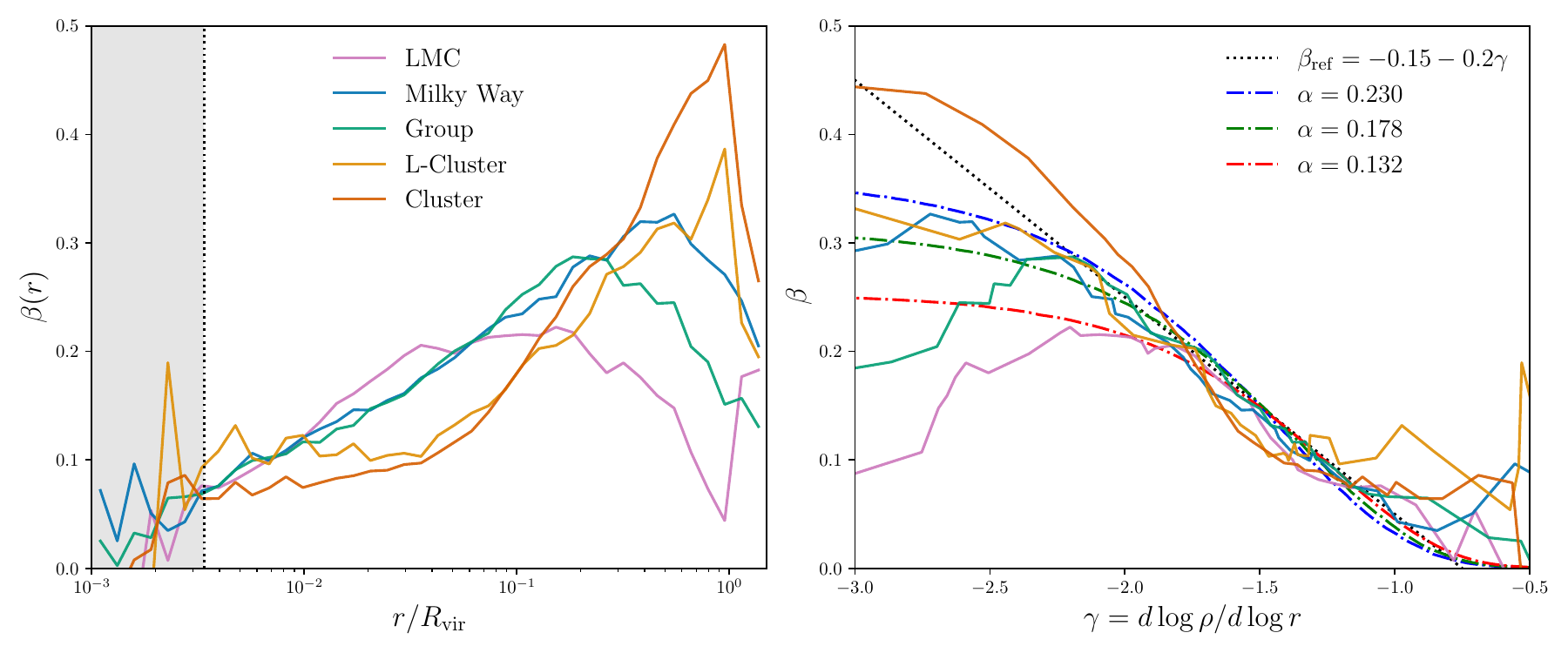}
    \caption{\textit{Left panel:} Velocity anisotropy profiles of host haloes in the \textit{Symphony} suites at $z=0$. The dotted vertical line marks the most conservative convergence radius across all suites, below which measurements may not be converged. \textit{Right panel:} Mean density slope--velocity anisotropy relation (Equation~\ref{eq:linear_beta_gamma}) for host haloes in \textit{Symphony} suites at $z=0$. The dotted line shows the linear relation proposed by \citet{2006NewA...11..333H}. The dot–dashed curves depict the model of \citet{2011MNRAS.415.3895L}, given by Equation~\ref{eq:beta_gamma}, assuming that $\gamma(r)$ follows an Einasto profile. The dot–dashed lines, from top to bottom, correspond to Einasto profiles with $\alpha = 0.230$, $0.178$, and $0.132$, respectively.
    }
\label{fig:velocity_anisotropy}
\end{figure*}

From Figure~\ref{fig:density_velocity}, it is apparent that the $r^2 \rho(r)$ and $\sigma(r)$ profiles follow very similar radial trends. This similarity implies the existence of a simple scaling relation between the density and velocity dispersion profiles, which partly motivates the construction of the PPSD (e.g., see the discussion in \citealt{Navarro_2009}).

\subsection{Velocity anisotropy profiles}

The left panel of Figure~\ref{fig:velocity_anisotropy} shows the mean velocity anisotropy (Equation~\ref{eq:vel_anisotropy}) profiles for all simulation suites at $z=0$, as a function of $r/R_{\mathrm{vir}}$. We find that host haloes in all suites have nearly isotropic velocity distributions in their central regions ($r/R_{\mathrm{vir}}\lesssim0.01$), typically with $\beta \lesssim 0.13$ at these radii. Beyond the central region, the L-Cluster and Cluster suites display a monotonic increase in $\beta(r)$ extending roughly out to the virial radius. This trend is consistent with previous studies of cluster-size haloes (e.g., \citealt{Ascasibar_2008, Lemze_2012, Wojtak_2013}). In these dynamically young systems, the velocity structure is dominated by strongly radial orbits characteristic of ongoing accretion, driving the anisotropy to rise continuously toward the halo outskirts. In contrast, the LMC, Milky Way, and Group suites' velocity anisotropy profiles exhibit a clear turnover at large radii: $\beta(r)$ increases with radius, indicating a transition from isotropic orbits in the innermost regions to more radially biased orbits near the scale radius, \textit{but then peaks and declines toward the outskirts}. The radius at which this transition occurs shows a systematic dependence on the suite mass: more massive hosts exhibit the transition at larger radii, and the peak value of $\beta$ at this location also tends to increase with suite mass. Similar knee-behavior has been reported in recent analyses of dark matter haloes in the TNG300 simulations \citep{2024ApJ...976..187H}.

We interpret this knee-behavior as marking the physical transition from a virialized, multi-stream interior region to an infall-dominated exterior region. To understand this feature, it is important to note that $\beta$ quantifies the anisotropy of the velocity dispersion (i.e., random motion) rather than its direction (i.e., bulk flow). In the virialized inner halo, the superposition of incoming and outgoing particles generates significant radial velocity dispersion, and thus higher $\beta$ values. In contrast, the region beyond the turnover is likely dominated by streams of accreting material. Although these particles have high radial velocities, their motion is coherent; this lack of variance suppresses the local radial dispersion, resulting in the observed dip in $\beta$. The systematic outward shift of this turnover radius with increasing host halo mass reflects the hierarchical assembly of haloes: the virialized regions of lower-mass, dynamically older haloes are more compact, whereas in massive, actively accreting systems, the infall regime extends closer to the virial boundary. We note that the systematic shift of this transition radius implies that a static, density-based definition of the virial radius does not align with haloes' dynamical boundaries, consistent with recent work on the splashback radius~\citep{Adhikari_2014,Diemer_2014,More_2015}.

\citet{2006NewA...11..333H} proposed that relaxed haloes follow a nearly universal linear relation between the density slope and the velocity anisotropy (the $\beta$–$\gamma$ relation), which can be expressed as
\begin{equation}
    \beta(\gamma) = \eta_1 - \eta_2 \gamma, \label{eq:linear_beta_gamma}
\end{equation}
where $-0.45 < \eta_1 < 0.05$ and $0.1 < \eta_2 < 0.35$. This empirical relation provides a closure condition for solving the Jeans equation in anisotropic systems. The right panel of Figure~\ref{fig:velocity_anisotropy} shows the mean $\beta$–$\gamma$ relations for all \textit{Symphony} suites, along with the \citet{2006NewA...11..333H} fit (dotted line). We find that the velocity anisotropy in the inner regions of \textit{Symphony} haloes, where $\gamma>-2$, can be well approximated by a universal parameter combination of $\eta_1 = -0.15$ and $\eta_2 = 0.2$ within $r_{-2}$, the radius at which the density slope equals $-2$. However, at larger radii where $-3 < \gamma < -2$, the profiles systematically deviate from the linear relation. This suggests that the linear $\beta$–$\gamma$ relation provides a good description of the inner halo regions but breaks down in the outskirts, where haloes are no longer fully relaxed.

\citet{2011MNRAS.415.3895L} further proposed that Equation~\ref{eq:linear_beta_gamma} can be replaced by a more flexible form,
\begin{equation}
    \beta(\gamma) = \frac{\beta_{\infty}}{2}\left[1 + \mathrm{erf}\!\left(\ln[(A\gamma)^2]\right)\right], \label{eq:beta_gamma}
\end{equation}
which better reproduces the behavior of the $\beta$–$\gamma$ relation in the outer halo regions. The dot–dashed curves in Figure~\ref{fig:velocity_anisotropy} illustrate the \citet{2011MNRAS.415.3895L} model, assuming that $\gamma(r)$ follows an Einasto profile \citep{1965TrAlm...5...87E}. The blue, green, and red dot–dashed lines correspond to $\alpha = 0.230$, $0.178$, and $0.132$, respectively. Thus, we find that the \citet{2011MNRAS.415.3895L} expression accurately describes the shapes of \textit{Symphony} $\beta(\gamma)$ profiles for $\gamma\gtrsim -2$, while we find systematically lower values of $\beta$ in the outskirts. This comparison indicates that \textit{Symphony} hosts are less relaxed than the \citet{2011MNRAS.415.3895L} haloes from which Equation~\ref{eq:beta_gamma} was derived; note that we do not apply a relaxation cut to our host samples, which could partially explain this difference.

\subsection{PPSD profiles}

\begin{figure*}
    \centering
    \includegraphics[width=1\linewidth]{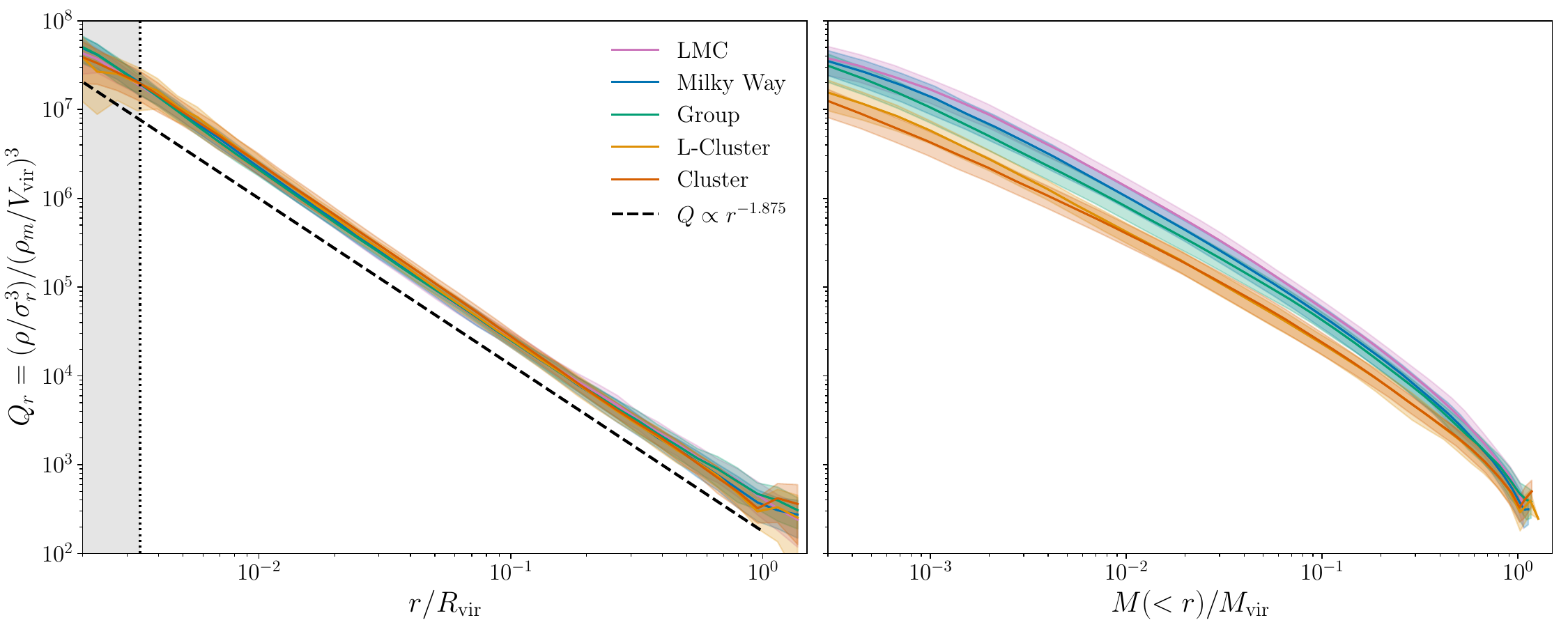}
    \caption{\textit{Left panel:} Mean radial PPSD $\rho/\sigma_{\mathrm{rad}}^3$ profiles of \textit{Symphony} host haloes, normalized by $\rho_m/V_{\rm vir}^3$, as a function of radius in units of $R_{\rm vir}$ for each suite. Shaded regions denote the 68\% halo-to-halo scatter. The dashed line represents the \citet{1985ApJS...58...39B} spherical secondary-infall similarity solution, $\rho/\sigma^3 \propto r^{-1.875}$. The dotted vertical line marks the most conservative convergence radius across suites. \textit{Right panel:} Same as the left panel, but shown as a function of enclosed mass within a certain radius (including substructures), normalized by the host virial mass $M_{\rm vir}$.}
    \label{fig:ppsd_profiles}
\end{figure*}

Figure~\ref{fig:ppsd_profiles} shows the mean measured radial PPSD profiles of simulated host haloes at $z=0$ in each suite in the same format as Figure~\ref{fig:density_velocity}. In the left panel of Figure~\ref{fig:ppsd_profiles}, the mean PPSD profiles from all simulation suites follow an almost identical power-law form, to remarkable precision, despite the fact that individual haloes exhibit diverse density and velocity structures even at fixed mass. At face value, the slope of this universal power-law strikingly resembles the self-similar solution $\rho/\sigma^3 \propto r^{-1.875}$ (dashed line), as predicted by spherical secondary infall models~\citep{1985ApJS...58...39B}. This apparent power-law behavior has also been identified in previous cosmological $N$-body simulations (e.g., \citealt{2001ApJ...563..483T,2005MNRAS.363.1057D,2010MNRAS.402...21N,Ludlow_2010,2011MNRAS.415.3895L}).

However, as we will demonstrate, PPSD profiles systematically deviate from a universal power law in a way that correlates with the host halo's dynamical state. One aspect of this non-universality is shown in the right panel of Figure~\ref{fig:ppsd_profiles}, where we plot the same mean PPSD profiles but as a function of enclosed mass, $M(<r)/M_{\rm vir}$ (i.e., using Lagrangian coordinates), rather than as a function of radius (i.e., using Eulerian coordinates). Unlike the radial profiles, the PPSD noticeably deviates from a power law when plotted as a function of enclosed mass \footnote{Mathematically, this deviation is partially expected due to the coordinate transformation: since the logarithmic slope of the mass profile, $\xi(r) \equiv d \ln M / d \ln r$, is not constant for realistic halos (e.g., NFW or Einasto profiles), a power law in radius, $Q(r) \propto r^\chi$, maps to $Q(M) \propto M^{\chi/\xi(r)}$, introducing curvature to the $Q(M)$ profile.}.
The PPSD mass profiles reveal a systematic deviation between the lower-mass (LMC, Milky-Way, and Group) and higher-mass (L-Cluster and Cluster) \textit{Symphony} suites. This separation likely arises due to significant differences in host halo density and velocity dispersion profiles between the lower-mass and higher-mass suites, which are apparent in Figure~\ref{fig:density_velocity}. \citet{Nadler_2023} show that these groups of \textit{Symphony} suites also differ in terms of their underlying host halo mass accretion histories (MAHs), which are known to correlate with halo density profiles (e.g., \citealt{2010arXiv1010.2539D,Ludlow_2013}).

We understand the difference between PPSD profiles as a function of radius and enclosed mass in Figure~\ref{fig:ppsd_profiles} as follows. When plotting PPSD profiles in Eulerian coordinates (i.e., as a function of radius), dynamical relaxation processes such as phase mixing and violent relaxation tend to smooth out profile differences to produce more universal behavior; in other words, a given radius generally contains contributions from multiple overlapping mass shells that have crossed and mixed during halo evolution. However, when plotting PPSD profiles in Lagrangian coordinates (i.e., as a function of enclosed mass), each mass shell retains the memory of the halo’s specific accretion and collapse history. This contrasts with averaging at a fixed Eulerian radius, which smooths out these halo-to-halo differences. Consequently, plotting in Lagrangian coordinates increases the PPSD profiles' sensitivity to differences in host MAHs.

\begin{figure}
    \centering
    \includegraphics[width=1\linewidth]{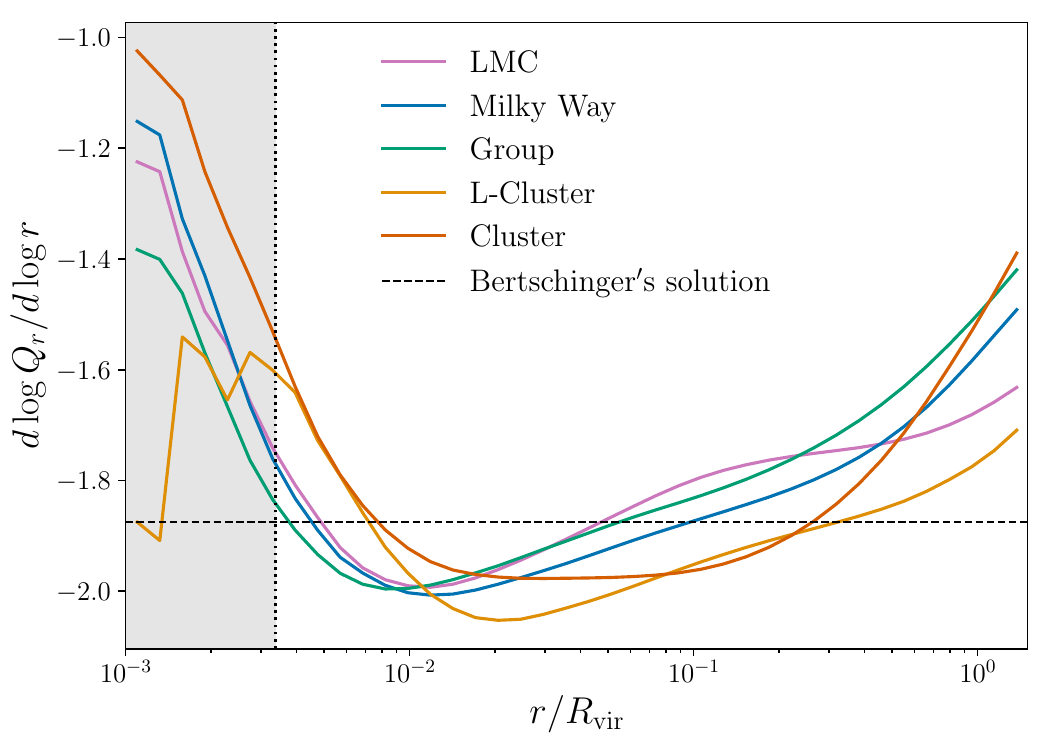}
    \caption{Mean logarithmic slope of PPSD profiles of host haloes in \textit{Symphony} suites at $z=0$. The horizontal dashed black line marks the universal relation $Q_r \propto r^{-1.875}$. The vertical dotted line marks the most conservative convergence radius across all \emph{Symphony} suites.}
    \label{fig:ppsd_slope}
\end{figure}

Figure~\ref{fig:ppsd_slope} shows the logarithmic slope of the PPSD profiles as a function of radius for each suite, compared to the single power-law behavior predicted by the secondary infall model of \cite{1985ApJS...58...39B}, $Q_r \propto r^{-1.875}$. We find that the PPSD profiles for each suite do not follow an exact power law, but tend to be shallower than $-1.875$ in the innermost and outermost regions, and steeper than $-1.875$ in the intermediate range, $0.01 R_{\rm vir} \lesssim r \lesssim 0.1 R_{\rm vir}$. Interestingly, these PPSD slope profiles are much largely independent of suite mass, which suggests that haloes of different masses have similar PPSD shapes.

The behavior in Figure~\ref{fig:ppsd_slope} is consistent with previous simulation results \citep[e.g.,][]{2009MNRAS.395.1225V, 2010MNRAS.402...21N, Ludlow_2010, 2011MNRAS.415.3895L} and indicates that PPSD profiles systematically deviate from a universal power law; as discussed above, this departure from a power law is even more severe as a function of enclosed mass. This non-universal behavior is consistent with the results of \citet{Nadler_2017}, who used fluid simulations to show that the PPSD does not follow a universal power law as a function of either enclosed mass or radius (also see \citealt{Arora_2020}). In the following section, we study the origins of this non-universal behavior.

\section{NON-UNIVERSALITY OF PPSD PROFILES}
\label{sec:non_universality}

\subsection{Deviation from Jeans Equilibrium}

\begin{figure*}
    \centering
    \includegraphics[width=1\linewidth]{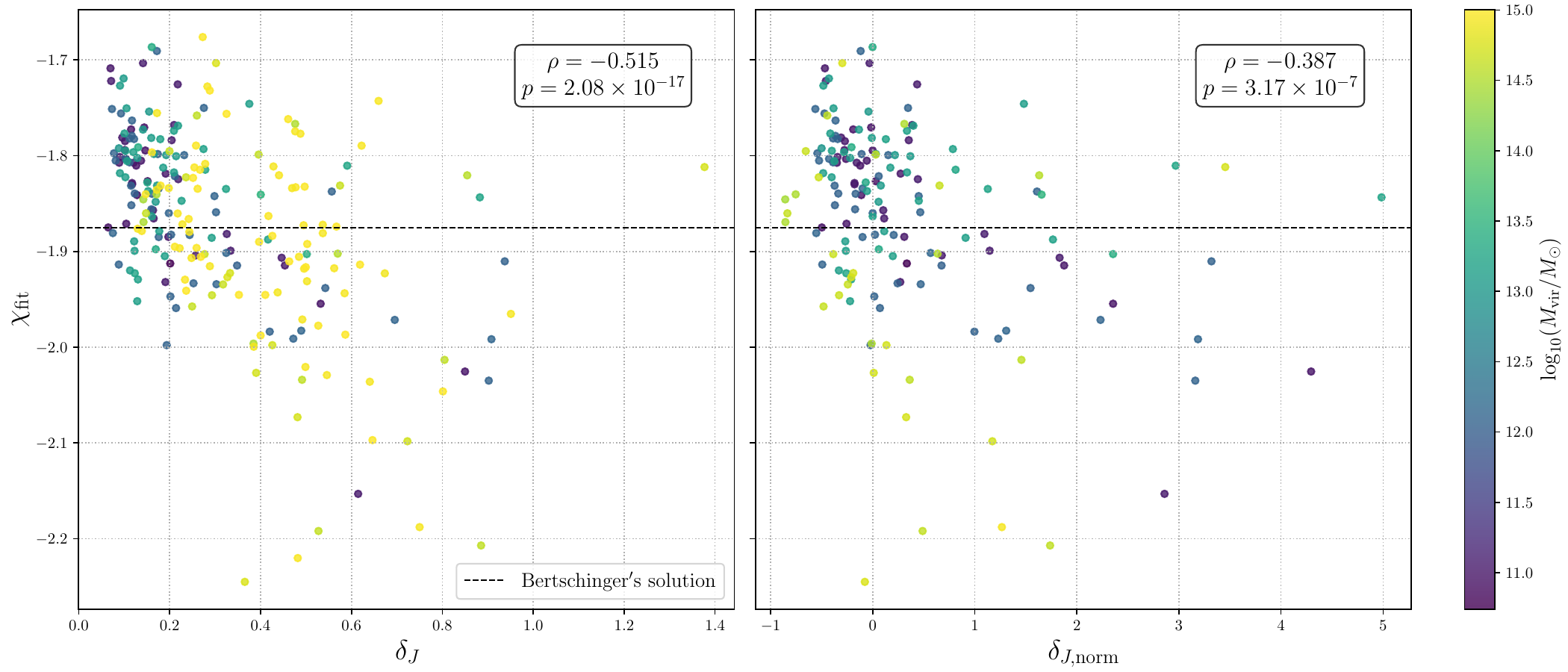}
    \caption{\textit{Left panel:} Best-fit logarithmic slope of radial PPSD profiles versus total Jeans deviation (Equation~\ref{eq:tot_jeans_dev}). The dashed line marks the universal relation $Q_r \propto r^{-1.875}$. The upper legend lists the Spearman correlation coefficient, which implies that hosts with larger $\delta_J$ systematically develop steeper PPSD slope. \textit{Right panel:} The same as the left panel but we normalize $\delta_J$ to $\delta_{J, \mathrm{norm}}$ (Equation~\ref{eq:normalize}) to remove the dependence on halo mass. Note that this panel excludes the Cluster suite as described in the text.}
    \label{fig:slope_vs_jeans}
\end{figure*}

Spherically symmetric dark matter haloes in equilibrium satisfy the Jeans equation:
\begin{equation}
    \frac{\mathrm{d}}{\mathrm{d}r}\!\left[\rho(r)\,\sigma_{\mathrm{rad}}^2(r)\right]
    + \frac{2\beta(r)}{r}\,\rho(r)\,\sigma_{\mathrm{rad}}^2(r)
    = -\rho(r)\,\frac{\mathrm{d}\Phi(r)}{\mathrm{d}r},
\end{equation}
where $\rho(r)$ is the density profile, $\sigma_{\mathrm{rad}}(r)$ the radial velocity dispersion, $\beta(r)$ the velocity anisotropy parameter, and $\Phi(r)$ the gravitational potential.  
In general, this equilibrium condition is not exactly satisfied by realistic haloes formed in $N$-body simulations. To quantify the degree to which haloes deviate from Jeans equilibrium, we define the \emph{Jeans deviation parameter} $\delta_J(r)$ as
\begin{equation}
\delta_J(r) \equiv
\left|
\frac{
\dfrac{\mathrm{d}}{\mathrm{d}r}\!\left[\rho(r)\,\sigma_{\mathrm{rad}}^2(r)\right]
+ \dfrac{2\beta(r)}{r}\,\rho(r)\,\sigma_{\mathrm{rad}}^2(r)
+ \rho(r)\,\dfrac{\mathrm{d}\Phi(r)}{\mathrm{d}r}
}{
\rho(r)\,\dfrac{\mathrm{d}\Phi(r)}{\mathrm{d}r}
}
\right|. \label{eq:jeans_dev}
\end{equation}
This dimensionless quantity measures the local imbalance between the pressure gradient and the gravitational force.
\footnote{Note that, because we use an absolute value in Equation~\ref{eq:jeans_dev}, our Jeans deviation parameter does not distinguish between ``pressure-driven'' versus ``gravitation-driven'' departures from Jeans equilibrium.}
By construction, a halo in perfect Jeans equilibrium at satisfies $\delta_J(r)=0$ at all radii, while larger values of $\delta_J(r)$ indicate stronger departures from equilibrium.

To assess the global level of disequilibrium within a halo, we define a \textit{total Jeans deviation parameter $\delta_J$} as:
\begin{equation}
\delta_J \equiv \int_{0}^{R_{\rm vir}} \delta_J(r)\,dr,  \label{eq:tot_jeans_dev}
\end{equation}
where $R_{\rm vir}$ is the virial radius.  
In Appendix~\ref{sec:jeans_deviaiton}, we demonstrate that this integrated quantity provides a consistent measure of the overall departure from equilibrium and correlates well with the commonly-used virial ratio $\eta = 2|K|/U$, where $K$ and $U$ respectively denote the total kinetic and potential energy.

\subsection{Jeans deviation drives PPSD non-universality}
\label{sec:jeans_dev}

We show the relation between the total Jeans deviation parameter $\delta_J$ and the best-fit PPSD slope in the left panel of Figure~\ref{fig:slope_vs_jeans}. The dashed line denotes the single power-law behavior predicted by the \cite{1985ApJS...58...39B} secondary infall model, $Q_r \propto r^{-1.875}$. To remove the mass dependence of $\delta_J$ among haloes within each simulation suite, the right panel shows PPSD slope versus the Jeans deviation normalized within each suite,
\begin{equation}
    \delta_{J, \mathrm{norm}} \equiv 
    \frac{\delta_J - \delta_{J, \mathrm{med}}}{\sigma_{\delta_J}}, \label{eq:normalize}
\end{equation}
where $\delta_{J, \mathrm{med}}$ and $\sigma_{\delta_J}$ denote the median and standard deviation of $\delta_J$ within the corresponding suite, respectively. Note that the Cluster suite is excluded in this figure, since its $\delta_{J,\text{norm}}$ distribution significantly differs from the other suites (see Appendix~\ref{sec:suites_consistency}). However, our results do not qualitatively change if the Cluster suite is included in the following analyses.

\begin{figure}
    \centering  
    \includegraphics[width=1\linewidth]{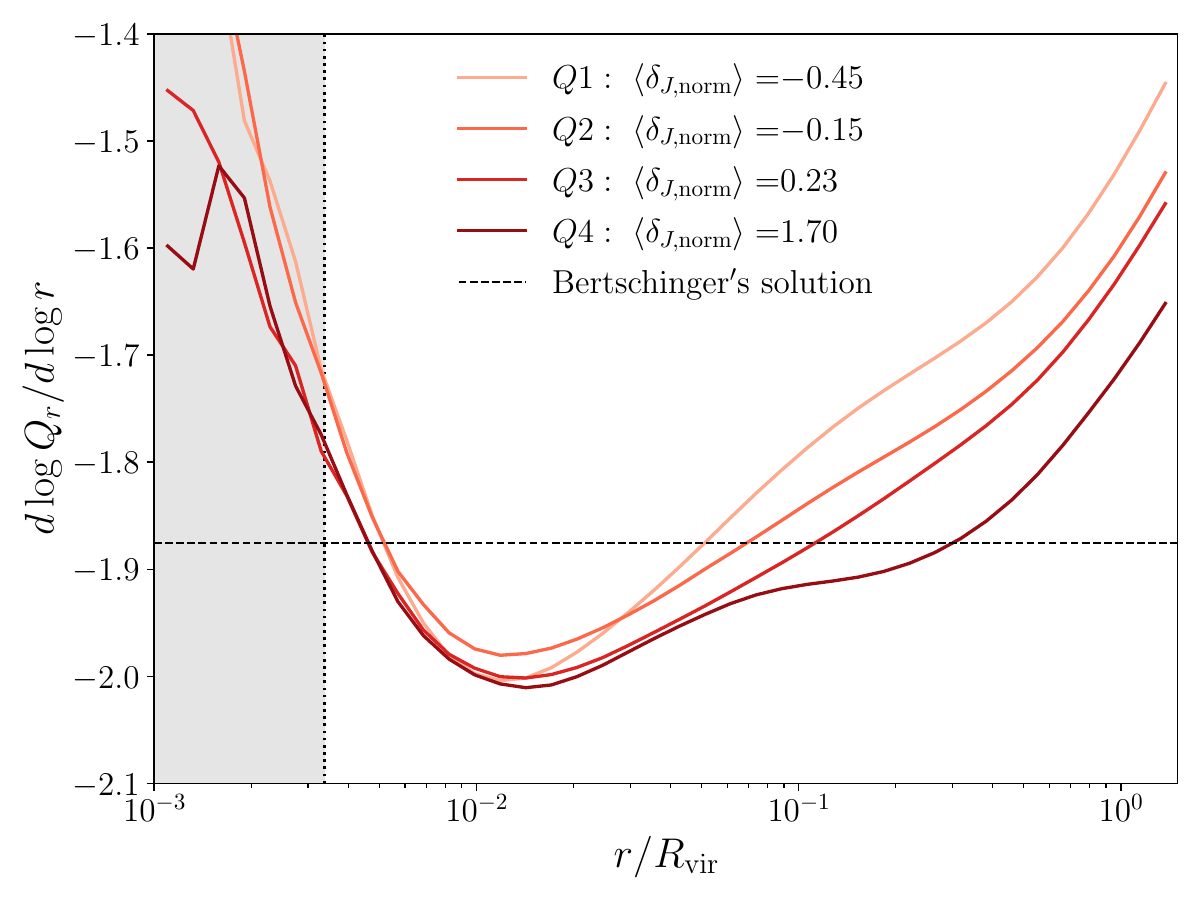}
    \caption{Mean radial profiles (solid lines) of the radial PPSD slope for \textit{Symphony} host haloes at $z=0$ (excluding the Cluster suite), split within each suite into quartiles of the normalized Jeans deviation, $\delta_{J,\rm norm}$. \emph{Q1} corresponds to the most relaxed haloes, while \emph{Q4} represents the most disturbed systems. The legend shows the mean normalized Jeans deviation for each quartile and the dashed line represents the universal relation $Q_r \propto r^{-1.875}$.}
    \label{fig:slope_quartiels_jeans}
\end{figure}

\begin{figure*}
    \centering
    \includegraphics[width=1\linewidth]{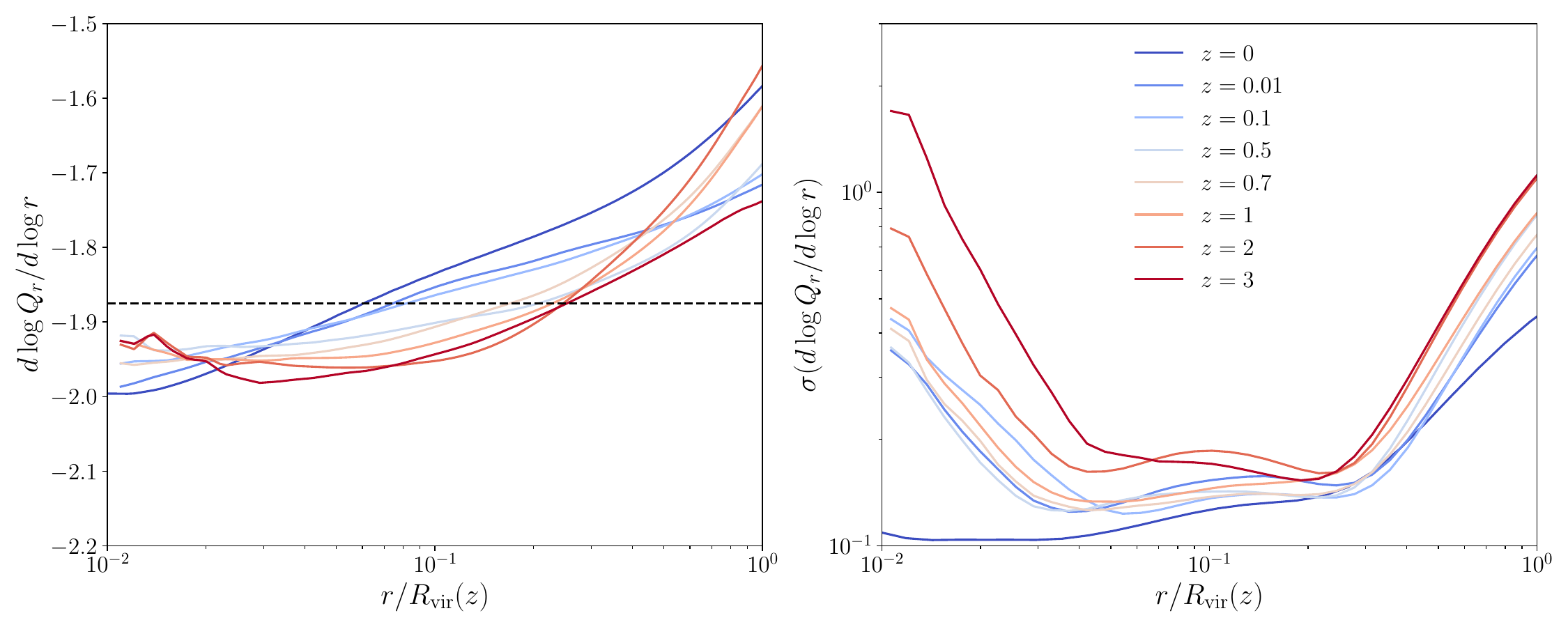}
    \caption{Evolution of the radial PPSD slope for host haloes in LMC, Milky-Way and Group suites. \textit{Left panel:} Mean logarithmic slope of $Q_r$ as a function of radius in units of $R_{\rm vir}(z)$ at different redshifts. The dashed line indicates the canonical $-1.875$ slope for reference. \textit{Right panel:} Radial profile of the standard deviation of the PPSD slope across all host haloes, which reveals the extent to which PPSD profiles differ from halo to halo as a function of radius and redshift.}
    \label{fig:slope_evolution}
\end{figure*}

Figure~\ref{fig:slope_vs_jeans} shows that haloes in different dynamical states can significantly deviate from the self-similar prediction. In particular, some haloes exhibit average PPSD slopes shallower than $-1.7$ or steeper than $-2.0$. More interestingly, we find a clear systematic trend: the farther a halo deviates from equilibrium (i.e., the larger $\delta_J$), the steeper its average PPSD slope becomes. The correlation between best-fit PPSD slope and normalized Jeans deviation is highly significant according to a Spearman correlation test ($\rho=-0.387$, $p=3.17\times 10^{-8}$). Thus, the strong correlation shown in the left panel of Figure~\ref{fig:slope_vs_jeans} is not purely driven by host halo mass. Note that these correlations remain strong and significant under jack-knife resampling, indicating that they are not driven by outlier hosts with large $\delta_J$ or $\delta_{J,\mathrm{norm}}$.

Based on this result, we suggest that the canonical relation $Q \propto r^{-1.875}$ observed in earlier $N$-body studies arises primarily from averaging haloes across a range of dynamical states, and that a halo's PPSD values strongly correlates with its departure from Jeans equilibrium. This echoes the findings of \citet{Nadler_2017}, which we will discuss below. We note that hints of this correlation between disequilibrium and PPSD slope have appeared in previous work. For example, \citet{Ludlow_2010} and \citet{2011MNRAS.415.3895L} suggested that deviations from a power law may reflect the presence of substructure or halo asphericity, both of which are known to indicate disequilibrium. Our finding that PPSD slopes correlate with departures from Jeans equilibrium provides a physical interpretation for these results.

\subsection{Evolution of non-universal PPSD profiles}
\label{sec:evolution}

Next, we consider how the PPSD slopes changes with radius and redshift. We divide the \textit{Symphony} host haloes (again excluding the Cluster suite) into quartiles according to their normalized Jeans deviation parameter. The mean PPSD slope profile for each quartile is shown in Figure~\ref{fig:slope_quartiels_jeans}. The first noticeable feature is that the more relaxed haloes systematically exhibit shallower PPSD slopes compared to those that are further from equilibrium, consistent with the trends discussed above.

In addition, the PPSD slope noticeably differs between radial regions. In the outer halo, the PPSD slopes display large scatter across different hosts, whereas in the inner region, the slopes are much less sensitive to the Jeans deviation parameter and tend to converge toward a nearly universal value of $\sim -2.0$. This behavior indicates that more dynamically relaxed regions within haloes tend to develop more self-similar and universal PPSD profiles.

\begin{figure}
    \centering
    \includegraphics[width=1\linewidth]{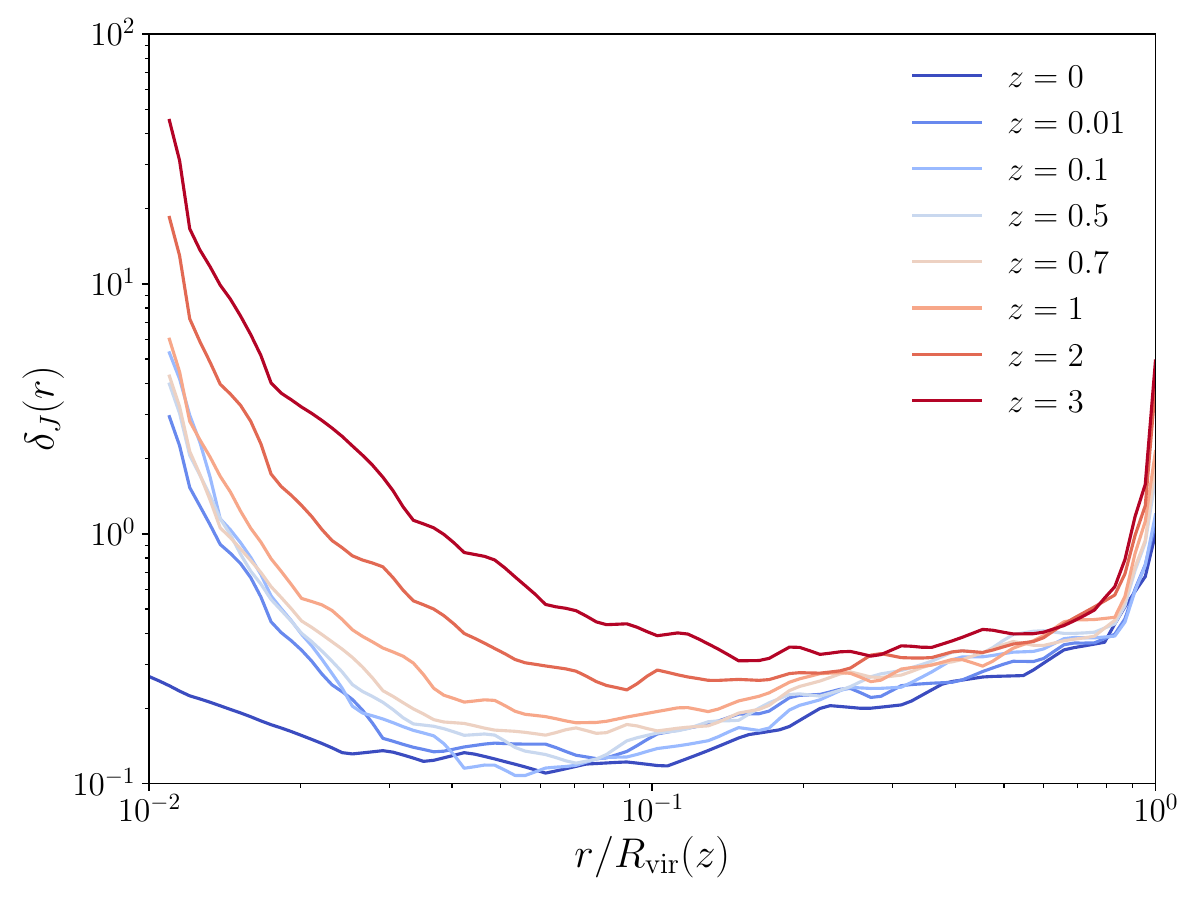}
    \caption{Average radial profiles of the Jeans deviation parameter, $\delta_J(r)$, for the LMC, Milky-Way, and Group suites shown in Figure~\ref{fig:slope_evolution}, as a function of redshift. Radii are scaled by each halo’s virial radius at the corresponding snapshot, and the profiles are averaged over all suites at each redshift.}
    \label{fig:jeans_evolution}
\end{figure}

\begin{figure*}
    \centering
    \includegraphics[width=1\linewidth]{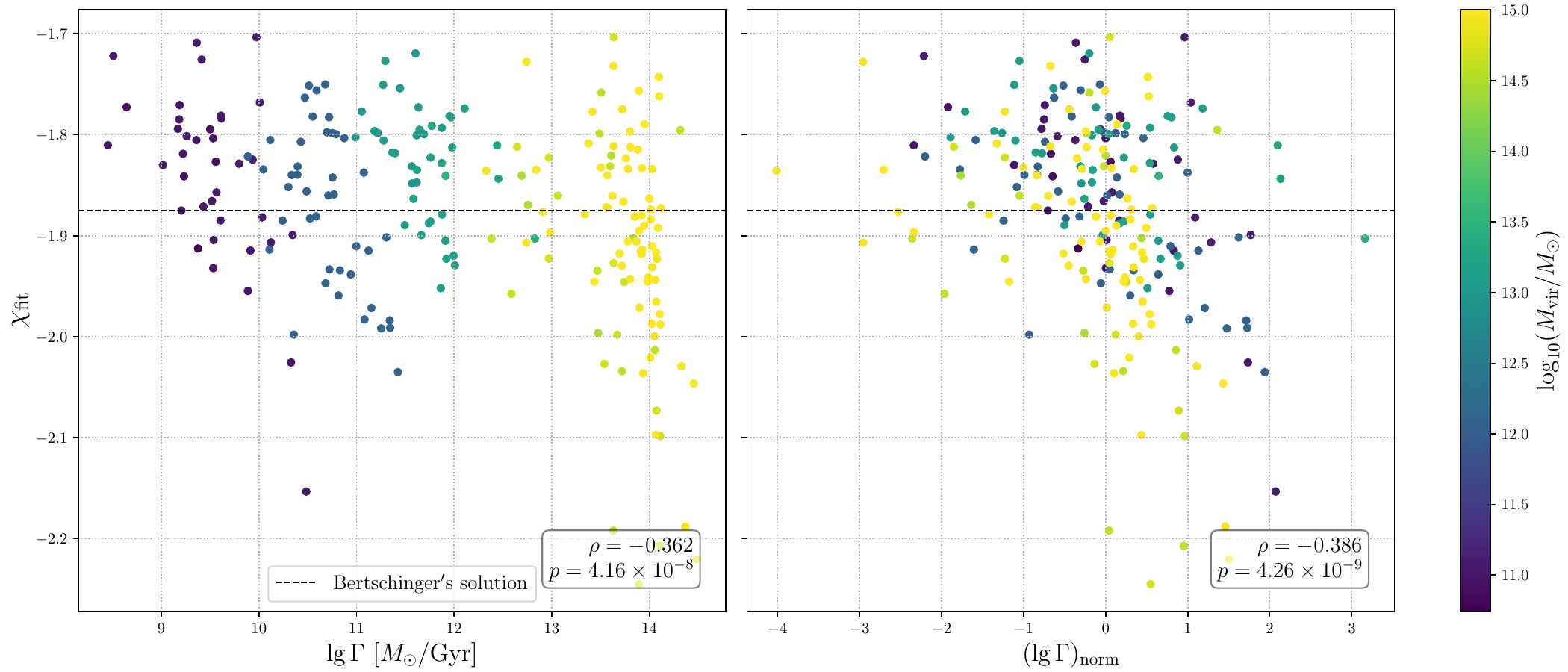}
    \caption{\textit{Left panel:} Best-fit logarithmic slope of radial PPSD profiles versus dynamical accretion rate at $z=0$ for all \textit{Symphony} hosts. The dashed line shows the universal relation $Q_r \propto r^{-1.875}$. \textit{Right panel:} The same as the left panel but with concentration normalized within each suite to remove the mass trend.}
    \label{fig:slope_vs_gamma}
\end{figure*}

\begin{figure*}
    \centering
    \includegraphics[width=1\linewidth]{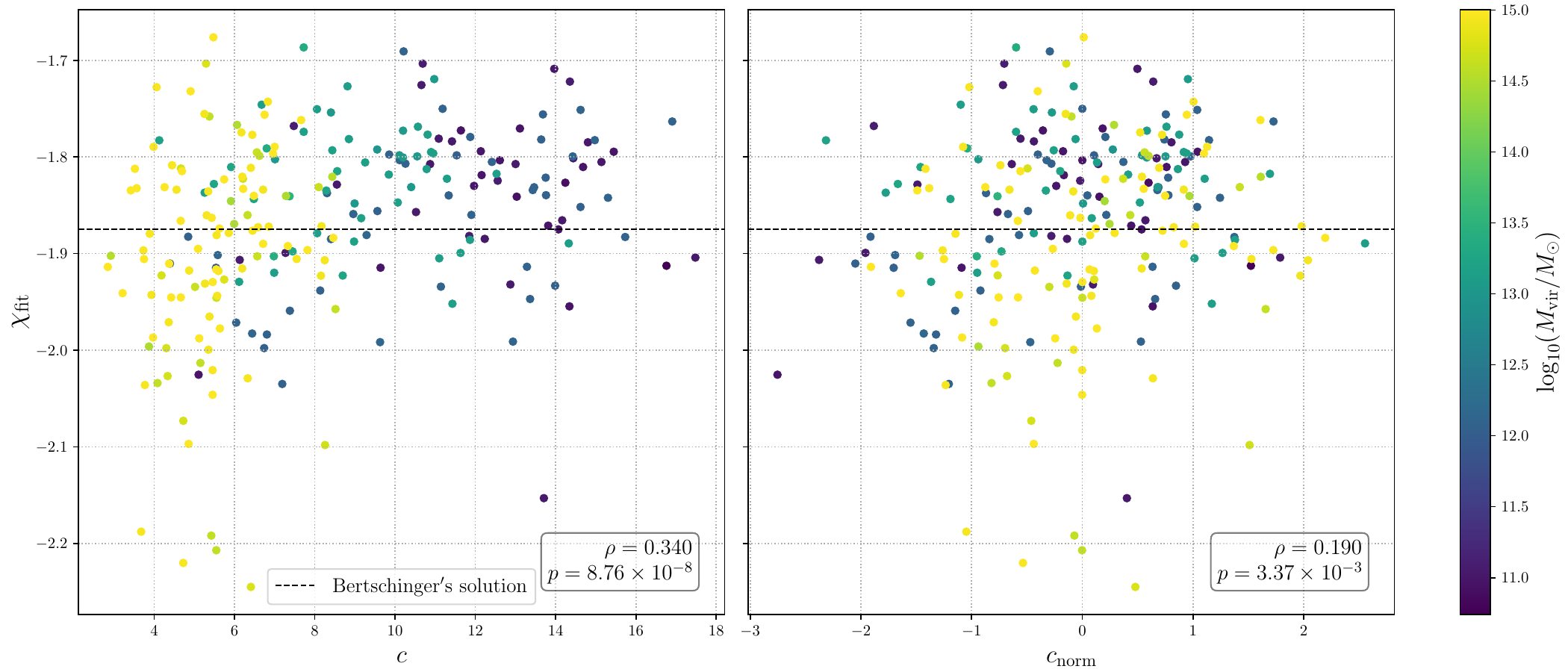}
    \caption{\textit{Left panel:} Best-fit logarithmic slope of radial PPSD profiles versus host halo's concentration at $z=0$ for all \textit{Symphony} suites. The dashed line represents the universal relation $Q_r \propto r^{-1.875}$. \textit{Right panel:} The same as the left panel but we normalize concentration within each suite to remove the mass trend.}
    \label{fig:slope_vs_cvir}
\end{figure*}

To clarify the origin of the PPSD shapes presented in Figure~\ref{fig:slope_quartiels_jeans} and to track their redshift evolution, Figure~\ref{fig:slope_evolution} shows the mean PPSD slope and the halo-to-halo scatter over time, as measured in comoving coordinates.\footnote{We exclude both the L-Cluster and Cluster suites for this analysis, since their mass accretion histories differ significantly from the other suites~\citep{Nadler_2023}; again, this choice does not qualitatively affect our results.} The PPSD retains an approximately power-law form, $Q(r)\propto r^{-1.875}$, with the same characteristic trend of being steeper in the inner halo and flatter in the outer halo, persisting back to at least $z=3$. This is qualitatively consistent with earlier findings (e.g., \citealt{Hoffman_2007}), who reported a PPSD power-law index near $-1.9$. However, with a substantially larger halo sample, we demonstrate robustly that the best-fit PPSD slope is steeper at higher redshifts (left panel of Figure~\ref{fig:slope_evolution}). Furthermore, the right panel of Figure~\ref{fig:slope_evolution} shows that the halo-to-halo scatter of the PPSD slope shrinks to small values in the inner region by $z=0$, while the halo-to-halo scatter is significantly larger in halo outskirts. These trends indicate that the inner PPSD is better approximated by a universal power law (albeit with an index somewhat steeper than $-1.875$), whereas PPSD profiles in the outer regions are less universal.

The behavior shown in Figure~\ref{fig:slope_evolution} can again be understood in terms of deviations from Jeans equilibrium. Figure~\ref{fig:jeans_evolution} shows the redshift evolution of the Jeans deviation parameter: haloes undergo rapid mass growth and frequent major mergers at high redshift, so their inner regions form and relax earlier while the outer regions continue to evolve. As a result, the radial and redshift-dependent scatter of the PPSD slope closely follows the evolution of the Jeans deviation, such that larger departures from equilibrium correspond to greater non-universality and systematically steeper PPSD slopes. This connection provides a natural explanation for why ensemble averages over haloes at different dynamical stages yield a power-law PPSD scaling, even though individual haloes may deviate substantially from this relation. Note that this argument only predicts that a power-law slope approximately emerges, but the specific \emph{value} of the power-law index does not have a deep physical significance in this picture. Thus, we interpret the value of the averaged PPSD slope (and the canonical $-1.875$ slope that is often reported) as a coincidence of averaging over haloes in different dynamical states.

\subsection{Correlations with haloes secondary properties}
\label{sec:secondary_properties}

Having demonstrated that the non-universality of the PPSD profile is primarily driven by deviations from Jeans equilibrium ($\delta_J$), we now investigate how the PPSD slope correlates with commonly measured secondary halo properties: the accretion rate and the virial concentration. Since $\delta_J$ is difficult to measure directly in observations, establishing correlations with these standard parameters provides a crucial link between our theoretical findings and observational diagnostics. To isolate the dependence on these properties from the halo mass trends, we utilize the normalized versions of these parameters within each simulation suite (e.g., $\Gamma_{\text{norm}}$ and $c_{\text{norm}}$), following the same normalization procedure described in Equation~\ref{eq:normalize}.

We first examine the accretion rate, which characterizes how rapidly the halo mass has grown at a specific redshift. We define this rate over the last dynamical time, $t_{\mathrm{dyn}}$, as:
\begin{equation}
\Gamma \equiv \frac{M_{\mathrm{vir}}(t_0) - M_{\mathrm{vir}}(t_0 - t_{\mathrm{dyn}})}{t_{\mathrm{dyn}}}.
\end{equation}
Figure~\ref{fig:slope_vs_gamma} illustrates the relationship between the best-fit PPSD slope and the dynamical accretion rate measured at $z=0$. We observe a significant systematic trend: haloes experiencing more intense recent accretion possess systematically steeper PPSD slopes. This correlation is robust across all suites, with a Spearman correlation coefficient of $\rho = -0.386$ ($p = 4.26 \times 10^{-9}$) for the normalized accretion rate.

This strong correlation reinforces the physical connection between the PPSD slope and the halo's dynamical state. Rapidly accreting haloes experience greater dynamical disturbances, driving them further from Jeans equilibrium. As demonstrated in Section~\ref{sec:jeans_dev}, this deviation leads to steeper PPSD slopes. Consequently, the shape of the PPSD profile serves as an indicator of the ongoing relaxation process initiated by mass assembly events.

Next, we examine the virial concentration, $c \equiv R_{\mathrm{vir}}/r_s$. Here, $r_s$ represents the characteristic scale radius obtained by fitting the halo density profile to the NFW formula \citep{Navarro_1997}. Unlike the accretion rate, which traces recent dynamical activity, concentration typically reflects the halo's earlier formation epoch \citep{Wechsler_2002}. Figure~\ref{fig:slope_vs_cvir} shows the relationship between the PPSD slope and concentration. In contrast to the accretion rate, the correlation here is notably weaker. While there is a slight tendency for more concentrated (typically earlier-formed) haloes to display shallower slopes, the scatter is large, and the correlation coefficient for the normalized concentration is only $\rho = 0.19$ ($p = 3.37 \times 10^{-3}$).

This result provides an interesting comparison to previous work. For example, \citet{Faltenbacher_2007} investigated the entropy profiles (where $K \propto Q^{-2/3}$) of dark matter haloes in cosmological smoothed particle hydrodynamics simulations and found that halo entropy profile shapes were effectively independent of concentration. Our analysis, utilizing the higher-resolution \textit{Symphony} zoom-in suites, confirms that this dependence is indeed weak. However, we detect a subtle secondary trend, such that higher concentration haloes exhibit slightly shallower PPSD slopes. Thus, we argue that the PPSD profile contains distinct dynamical information rather than merely reflecting the density profile shape. Two haloes with identical concentrations can have significantly different PPSD slopes depending on their recent accretion activity and degree of dynamical equilibrium.

\section{COMPARISON TO THE one-dimensional FLUID COLLAPSE SIMULATION}
\label{sec:fluid_collapse}

For a monatomic ideal gas, the local specific entropy (i.e., entropy per particle) can be expressed as
\begin{equation}
    s_{\rm gas}(r) = - k_{\rm B} \ln K_{\rm gas}^{3/2}(r) + \mathrm{constant},
\end{equation}
where $k_{\rm B}$ is the Boltzmann constant.
The quantity $K_{\rm gas}$ denotes the conventional “entropy” of the intracluster medium (ICM) widely used in X-ray observations, defined as
\begin{equation}
    K_{\rm gas} \equiv \frac{3 k_{\rm B} T_{\rm g}}{\mu m_{\rm p}} \rho_{\rm g}^{-2/3}
    = \sigma_{\rm g}^{2}\, \rho_{\rm g}^{-2/3},
    \label{eq:entropy}
\end{equation}
where $T_{\rm g}$, $\rho_{\rm g}$, and $\sigma_{\rm g}$ are the gas temperature, density, and velocity dispersion, respectively, and $\mu m_{\rm p}$ is the mean particle mass (\citealt{Dalcanton_2001}).

Equation~\ref{eq:entropy} provides a natural analogue for defining the thermodynamic entropy of a collisionless dark matter (DM) system~\citep{Faltenbacher_2007},
\begin{equation}
    K_{\rm DM} \equiv \sigma_{\rm DM}^{2} \rho_{\rm DM}^{-2/3} = Q^{-2/3}.
\end{equation}
Both numerical simulations and X-ray observations of galaxy clusters have shown that the gas entropy and dark matter PPSD respectively follow a radial scaling of $K \propto r^{1.2}$ and $Q \propto r^{-1.8}$, supporting this correspondence (e.g., \citealt{Faltenbacher_2007, 2009ApJS..182...12C,2019MNRAS.482.1043C,Marini_2020}).\footnote{For ICM gas in observed galaxy clusters, the entropy profile can be flattened in the inner regions due to AGN feedback and radiative cooling, but follows a power law at large radii where such effects become negligible (e.g., \citealt{2009ApJS..182...12C}).}
This empirical similarity suggests that the thermodynamic structuring of hot gas and the dynamical relaxation of dark matter are governed by analogous physical principles.

\begin{figure}
    \centering
    \includegraphics[width=1\linewidth]{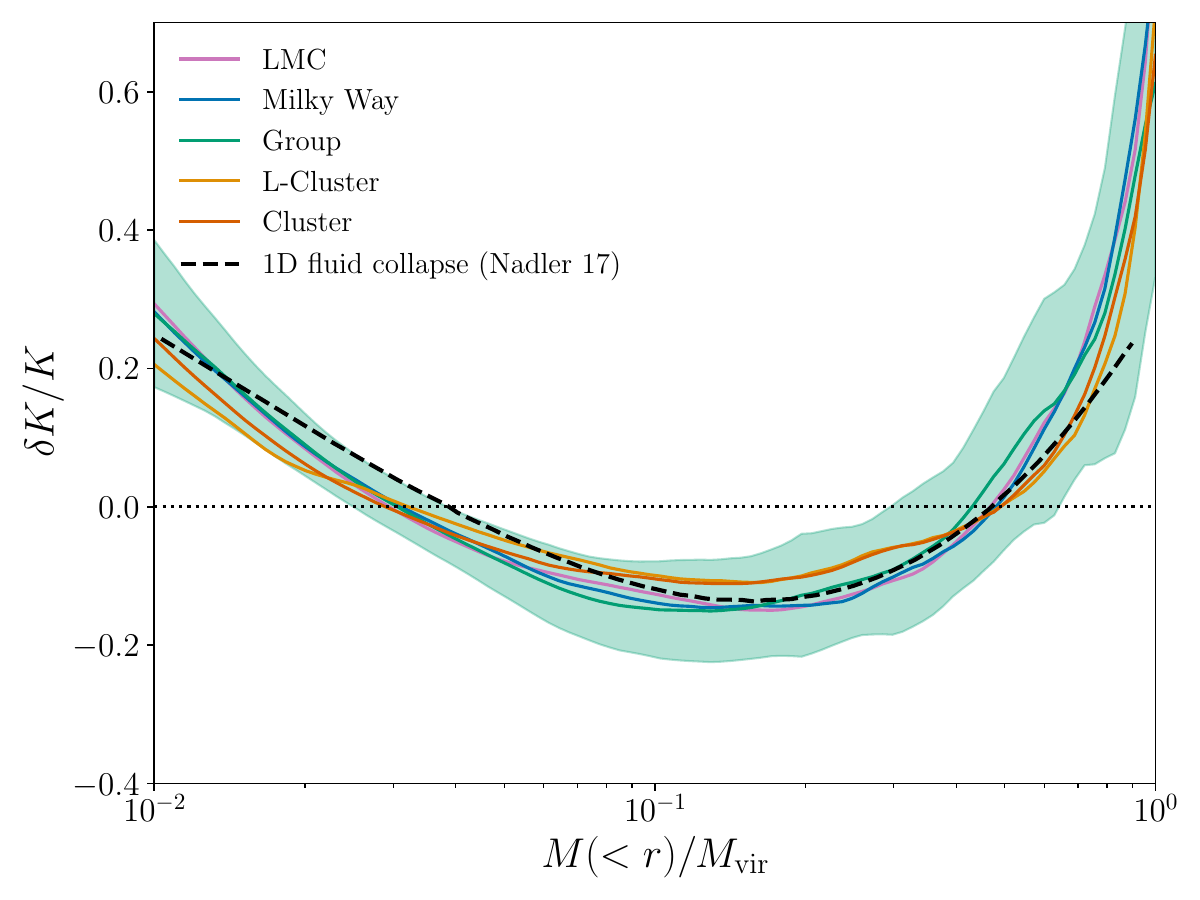}
    \caption{Stacked residuals of the entropy profile of \textit{Symphony} host haloes within each suite relative to power-law fits, compared with the one-dimensional realistic fluid collapse model of \citet{Nadler_2017}. The residuals, $(K_{\rm fit} - K)/K$ with $K = Q_r^{2/3}$, are shown as a function of enclosed mass $M(<r)/M_{\rm vir}$ for all \textit{Symphony} host haloes. The shaded band denotes the $1\sigma$ halo-to-halo scatter of the Group suite and the black dashed curve is the result of \citet{Nadler_2017}. The horizontal dotted line corresponds to a perfect power-law entropy profile.}
    \label{fig:entropy_residual}
\end{figure}

Following this connection, \citet{Nadler_2017} employed a fluid approximation to investigate deviations from the idealized power-law PPSD (or entropy) profile. Figure~\ref{fig:entropy_residual} compares our simulation results with their one-dimensional hydrodynamic simulation predictions that employ a realistic halo mass accretion history (i.e., the blue line in Figure 10 of \citealt{Nadler_2017}). We compare as a function of enclosed mass rather than radius because the \citet{Nadler_2017} radial profiles are normalized to the shock radius in their hydrodynamic simulations, which is not well defined in our N-body haloes. 

The residuals of our radial entropy measurement agree remarkably well with the \citet{Nadler_2017} prediction, indicating that one-dimensional fluid simulations accurately reproduce entropy (and thus PPSD) power-law deviations found in our three-dimensional cosmological N-body zoom-in simulations. This correspondence implies that the shape of the PPSD and its deviation from a power law are primarily determined by departures from Jeans (or, in a fluid context, hydrostatic) equilibrium, as encoded in halo assembly history, consistent with the predictions in \citet{Nadler_2017}.

These results demonstrate that the one-dimensional self-similar fluid collapse model accurately captures the \textit{mean} behavior of PPSD profiles in cosmological simulations. This agreement holds not only for the ensemble average across all hosts, but also for the average of each individual simulation suite, suggesting that the correspondence is robust across mass scales. While three-dimensional dynamical effects including velocity anisotropy and filamentary accretion likely drive the significant halo-to-halo scatter observed in Figure~\ref{fig:entropy_residual}, they do not systematically bias the mean PPSD profile away from the one-dimensional fluid prediction. This interpretation is consistent with the close similarity between the mean total and radial PPSD profiles in our simulations (Appendix~\ref{sec:total_ppsd}).
Furthermore, these results suggest that the radial orbit instability (ROI) does not significantly affect the PPSD, consistent with the arguments in \citet{Nadler_2017}, despite the importance of the ROI in shaping haloes' central density and velocity disperion profiles (e.g., \citealt{MacMillan_2006}).

\section{Summary and Discussion}
\label{sec:summary}

\subsection{Main Findings}

We analyzed the density, velocity dispersion, velocity anisotropy, and pseudo–phase-space density (PPSD) profiles of 246 host haloes spanning five mass decades from the \textit{Symphony} simulation suite. We investigated how the PPSD profiles deviate from the previously reported universal power law relation and explored how these deviations correlate with the degree of dynamical equilibrium, mass accretion history, and other secondary halo properties.

Our main findings are as follows:
\begin{itemize}
    \item \textit{PPSD profiles do not exactly follow a universal power law.}
    In particular, \textit{Symphony} hosts' PPSD profiles systematically deviate from the canonical $Q \propto r^{-1.875}$ power law. Across all suites, the logarithmic slope of $Q(r)$ is steeper than~$-1.875$ in the inner regions ($r/R_{\mathrm{vir}}\lesssim 10^{-1}$) and shallower at larger radii (Figure~\ref{fig:ppsd_slope}).
\end{itemize}

\begin{itemize}
    \item \textit{haloes farther from Jeans equilibrium exhibit steeper average PPSD slopes.} 
    haloes that deviate more strongly from Jeans equilibrium, as quantified by our Jeans deviation parameter (Equation~\ref{eq:jeans_dev}), have significantly steeper average PPSD slopes (Figure~\ref{fig:slope_vs_jeans}). This correlation remains highly significant, even after controlling for halo mass and performing jackknife resampling.
\end{itemize}

\begin{itemize}
    \item \textit{Halo mass assembly histories shape PPSD profiles.}
    The average PPSD slope is steep (close to $-2.0$) in the inner halo ($r/R_{\mathrm{vir}}\lesssim 10^{-1}$) and becomes increasingly stable toward lower redshifts. At larger radii, however, the slope exhibits substantial halo-to-halo scatter that persists until the present day (Figure~\ref{fig:slope_evolution}). This behavior reflects halo assembly: rapid accretion and mergers at early times drive strong departures from Jeans equilibrium and produce the large variability in outer PPSD slopes, whereas subsequent relaxation leads to the convergence toward a more universal inner profile.
    \end{itemize}

\begin{itemize}
    \item \textit{PPSD profiles depend on halo concentration and accretion rate.}
    The best-fit average PPSD slopes show a weak but systematic correlation with concentration and a much stronger correlation with recent accretion rate: more concentrated (and thus dynamically older) haloes tend to have slightly shallower slopes, whereas rapidly accreting haloes exhibit significantly steeper slopes (Figures~\ref{fig:slope_vs_cvir}–\ref{fig:slope_vs_gamma}). These trends persist after controlling for host halo mass.
    \end{itemize}

\begin{itemize}
    \item \textit{one-dimensional self-similar fluid collapse model largely recovers the shape of PPSD profiles in N-body simulations.}
    The PPSD profiles measured in our simulations are in remarkable agreement with predictions from the \citet{Nadler_2017} one-dimensional fluid model (Figure~\ref{fig:entropy_residual}). Deviations from the ideal power-law PPSD can therefore be explained by departures from Jeans equilibrium, which are ultimately set by halo assembly history alone.
\end{itemize}

Thus, we argue that a power-law PPSD approximately emerges from ensemble averaging over haloes in different dynamical states, and that PPSD profiles deviate from a universal power-law relation because they encode haloes' diverse mass assembly histories and dynamical states. In this picture, the PPSD is primarily governed by quasi-hydrostatic relaxation associated with mass assembly. Three-dimensional effects, such as velocity anisotropy and filamentary accretion, appear to be relatively unimportant in setting the PPSD profiles compared to the halo's dynamical state.

\subsection{Outlook and Future Work}

These results build on previous studies in several ways. For example, \citet{Nadler_2017} found that the PPSD deviates from a universal power-law relation because of deviations from hydrostatic equilibrium in their one-dimensional hydrodynamic simulations. Our results confirm this prediction in a three-dimensional cosmological setting without relying on the fluid approximation. Moreover, the striking similarity between our PPSD profile and the (appropriately transformed) \citet{Nadler_2017} gas entropy profile (Figure~\ref{fig:entropy_residual}) suggests that the PPSD is primarily determined by halo mass assembly history, clarifying the interpretation of the \citet{Nadler_2017} results. Meanwhile, \citet{Arora_2020} argued that a power-law PPSD is not a fundamental outcome of gravitational evolution based on the structure of the Jeans equation, even for equilibrium haloes. Our results strengthen this conclusion for non-equilibrium haloes.

A critical next step is to investigate the impact of baryonic physics on the PPSD correlations we have found using full hydrodynamic simulations. As noted earlier, our dark matter-only analysis isolates gravitational dynamics, but baryons can significantly reshape inner halo structure. It remains to be seen whether the correlations we establish here (e.g., between PPSD slope and dynamical state) persist in the presence of dissipative processes. In addition, hydrodynamic simulations are essential for making direct comparisons with observations, as the PPSD is often inferred from the thermodynamic properties of intracluster gas \citep{2009ApJ...692..174L, 2010A&A...511A..85P, Biviano_2023}. By analyzing the phase-space structure of both dark matter and baryons in a hydrodynamic setting, future work can bridge the gap between our theoretical predictions and galaxy cluster observations.

Our results are based on \textit{Symphony} CDM simulations with fixed cosmological parameters within each suite, raising the question of how the PPSD behaves in other cosmologies. For example, \citet{Brown_2020} found that PPSD is sensitive to the shape of the primordial matter power spectrum and that the apparent universality of PPSD profiles may result from the fairly narrow range of initial conditions used in most cosmological simulations. Our key result that halo mass assembly history determines the PPSD implies that variations in the PPSD are determined by the impact of these initial conditions on halo mass assembly histories. This is plausible, since mass assembly histories reflect the initial conditions for structure formation (e.g., \citealt{Lu_2006}). A dedicated study is needed to test whether our prediction is consistent with the \citet{Brown_2020} results.

In a similar vein, our framework can be used to understand the PPSD in dark matter models beyond CDM. For example, in models that alter the initial conditions for structure formation like warm dark matter (WDM), we predict that changes to the PPSD relative to CDM will be determined by the impact of power spectrum suppression on halo growth histories. This is a timely area for future work given previous studies on the WDM PPSD (e.g., \citealt{Dalcanton_2001,Maccio_2012,Shao_2013}) as well as the large suites of WDM simulations now available (e.g., \citealt{Rose_2024,Nadler_2025_COZMIC}). Non-gravitational dark matter physics can also alter the PPSD. For example, self-interacting dark matter (SIDM) models produce central cores in density and velocity dispersion profiles, leading to a cored inner PPSD (e.g., \citealt{Rocha_2012,Chua_2020}). It will also be interesting to revisit the PPSD in modern SIDM simulations, which adopt high-amplitude, velocity-dependent cross sections and produce both core formation and collapse (e.g., \citealt{Nadler_2025_Concerto}). 

Finally, we note that studying the PPSD and its evolution in subhaloes is particularly interesting in light of our results, since subhaloes deviate even more substantially from Jeans equilibrium than isolated haloes during orbital evolution. For example, \citet{2009MNRAS.395.1225V} found that the amplitude of the coarse-grained distribution function $f$ (and thus of the PPSD) is orders-of-magnitude higher at the centers of subhaloes compared to the main halo because this relatively unmixed material better captures the high primordial dark matter phase-space density at high redshifts. This high amplitude of the PPSD in subhaloes is hinted at in Figure~\ref{fig:ppsd_visual} and raises the question of how strongly our hosts' PPSD profiles are impacted by subhaloes when including all particles (rather than only particles bound to the host), similar to previous work for halo density profiles (e.g., \citealt{Fielder_2020}). We expect that analyses of higher-resolution \emph{Symphony} resimulations combined with recent analytic techniques for evolving subhalo phase-space distributions (e.g., \citealt{Drakos_2017}) will shed light on these issues.

\section*{Acknowledgements}

We thank Fangzhou Jiang, Phil Mansfield, and Jorge Pe\~narrubia for helpful discussions related to this work. This work used data from the \textit{Symphony} suite of simulations, which were supported by the Kavli Institute for Particle Astrophysics and Cosmology at Stanford University and SLAC National Accelerator Laboratory, and by the U.S. Department of Energy under contract number DE-AC02-76SF00515 to SLAC National Accelerator Laboratory. S.P.O.\ acknowledges National Science Foundation (grant number AST-240752) for support. S.J. acknowledges National Science Foundation of China (grant number 12522301) for support.

\section*{Data Availability}

Our analysis code is publicly available at \url{https://github.com/Bocheng-Feng/Symphony-PPSD}. All \textit{Symphony} data we used is hosted at \url{http://web.stanford.edu/group/gfc/gfcsims}.


\bibliographystyle{mnras}
\bibliography{references}



\appendix

\twocolumn

\section{Alternative PPSD Measurements}

\subsection{Total versus radial PPSD}
\label{sec:total_ppsd}

\begin{figure*}
    \centering
    \includegraphics[width=1\linewidth]{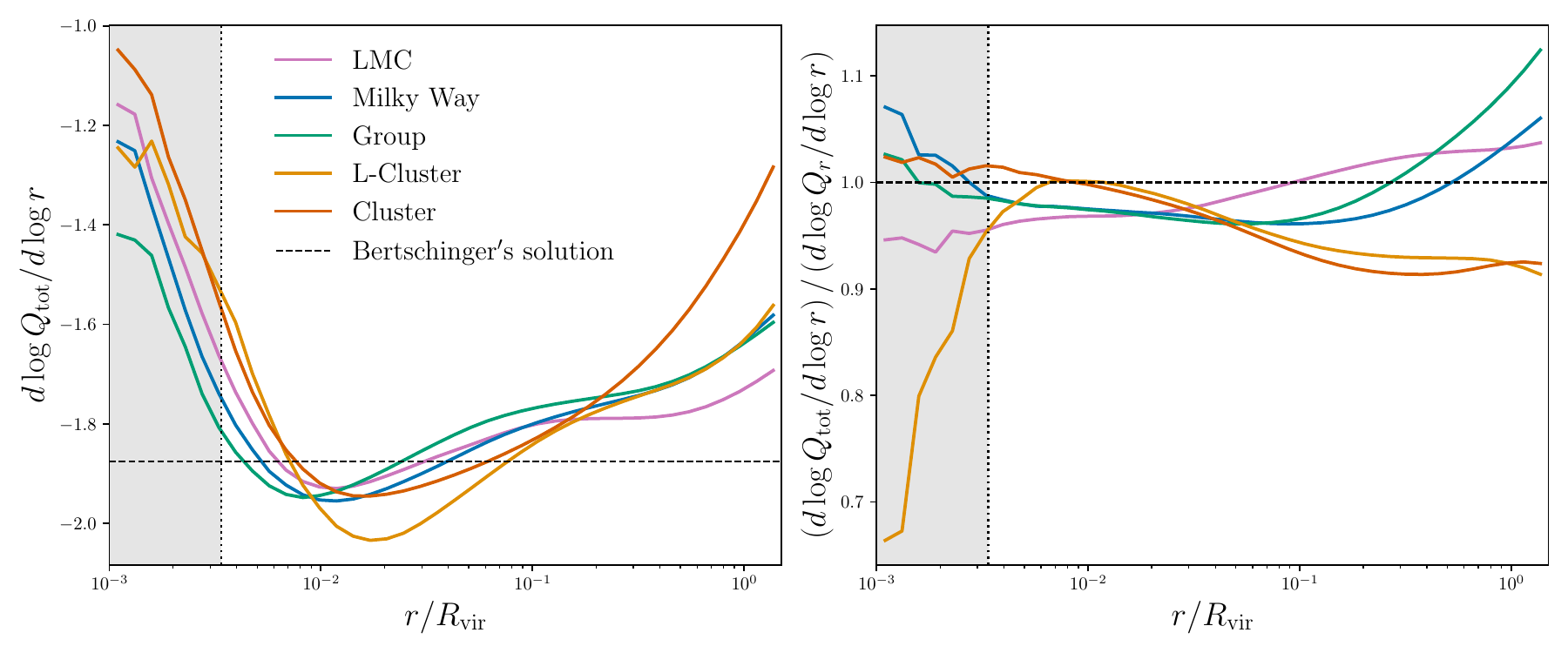}
    \caption{\textit{Left panel:} Mean logarithmic slope of radial PPSD profiles of host haloes in \textit{Symphony}
suites at $z=0$. The dotted vertical line marks the most conservative convergence radius across all suites. \textit{Right panel:} The ratio between total and radial PPSD slope profiles.}
    \label{fig:Qtot_vs_Qr}
\end{figure*}

The total and radial pseudo–phase–space densities (PPSDs) are related through the expression
\begin{equation}
    \frac{Q_{\mathrm{tot}}(r)}{Q_{r}(r)} = [3 - 2\beta(r)]^{-3/2},\label{eq:q_ratio}
\end{equation}
where $\beta(r)$ denotes the velocity anisotropy profile of the halo. 
Figure~\ref{fig:Qtot_vs_Qr} compares the total PPSD with the radial one. We find that their overall shapes and slopes exhibit very similar behavior across all suites and that their ratio is consistent with Equation~\ref{eq:q_ratio}. 

Thus, the arguments presented in the main text for the radial PPSD equally apply to the total PPSD. In particular, while the slope of the total PPSD tends to be slightly shallower than that of the radial PPSD, its correlations with the Jeans deviation and other halo properties remain robust.

\subsection{Dimensionful PPSD}
\label{sec:dimensional_ppsd}

\begin{figure*}
    \centering
    \includegraphics[width=1\linewidth]{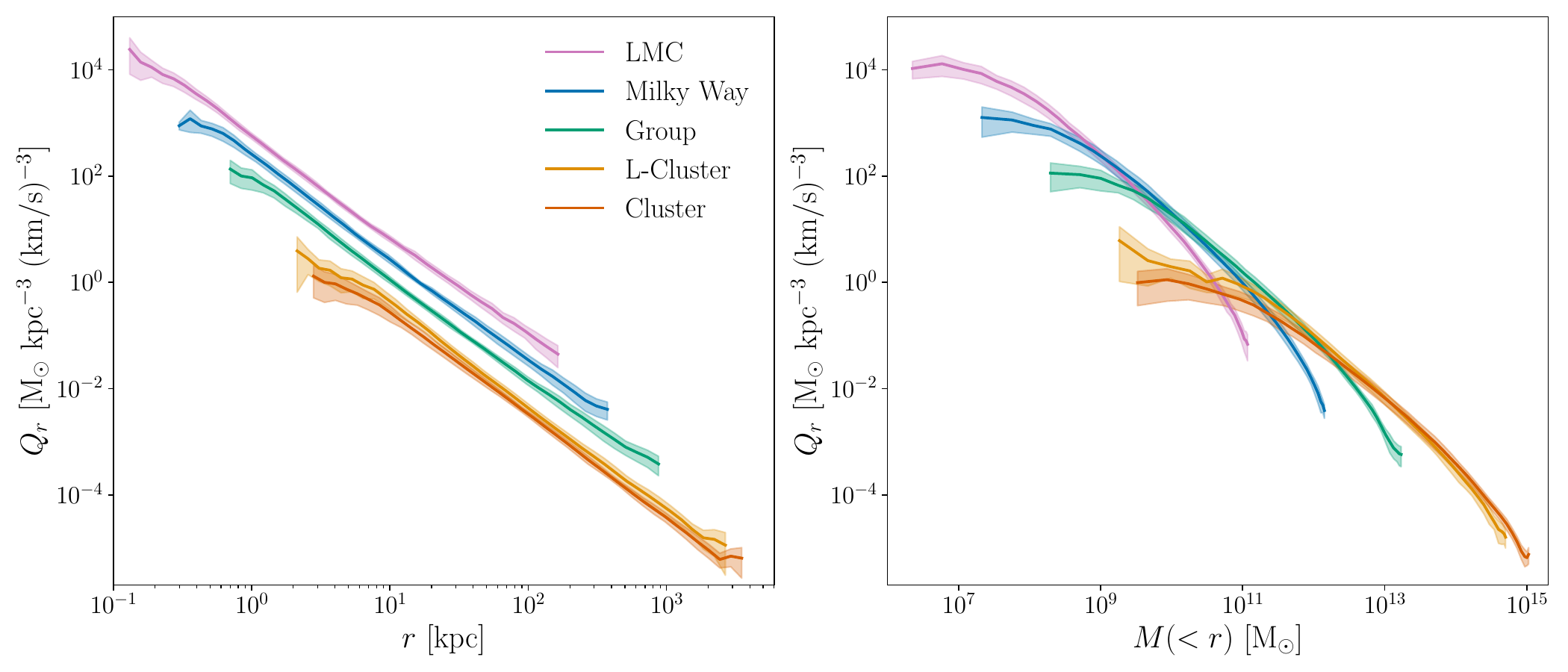}
    \caption{Same as Figure~\ref{fig:ppsd_profiles}, but showing the dimensionful radial PPSD as a function of dimensionful radius (left panel) and dimensionful nclosed mass (right panel, in units of $M_{\odot}$).}
    \label{fig:dimensionful_ppsd}
\end{figure*}

Figure~\ref{fig:dimensionful_ppsd} presents the mean radial PPSD profiles for each simulation suite in physical units of $M_{\odot}\,\mathrm{kpc}^{-3}\,(\mathrm{km\,s^{-1}})^{-3}$, shown as functions of radius (left panel, in units of $\mathrm{kpc}$) and enclosed mass (right panel, in units of $M_{\odot}$). We find that the amplitude $A$ in the power-law model $Q = A (r/R_{\mathrm{max}})^{-\chi}$ correlates with host halo mass.\footnote{In this relation, we normalize radii by $R_{\mathrm{max}}$, the radius from the center of each halo at which $V_{\mathrm{max}}$ is achieved, to remove residual mass dependence across our suites.} In particular, the amplitude of the best-fitting single power-law model follows a clear scaling relation:
\begin{equation}
    \lg{A} \approx (-1.20 \pm 0.01) \lg{M_{\rm vir}} + (13.54 \pm 0.19), 
    \label{eq:ppsd_amplitude}
\end{equation}
where $A$ represents the value of radial PPSD at $R_{\rm max}$, the radius where the circular velocity reaches its maximum and $M_{\rm vir}$ denotes the virial mass of the host halo, and uncertainties represent $1\sigma$ intervals about the best-fit parameters. This scaling is consistent with the trend proposed by \citet{Dalcanton_2001}, who argued that systems assembled through hierarchical structure formation obey $Q \sim M^{-1}$. 

This scaling arises because systems that assemble at roughly constant spatial density must undergo proportional increases in both size and velocity dispersion to maintain virial equilibrium. This leads phase-space volume to increase more rapidly than total mass, resulting in a characteristic decline of the coarse-grained phase-space density with increasing mass, i.e., $Q \propto M^{-1}$. The relation thus represents the slowest possible decrease of $Q$ with mass, corresponding to the most quiescent merger histories.

\section{Jeans deviation versus virial ratio}
\label{sec:jeans_deviaiton}

\begin{figure*}
    \centering
    \begin{minipage}{0.48\textwidth}
        \centering
        \includegraphics[width=\linewidth]{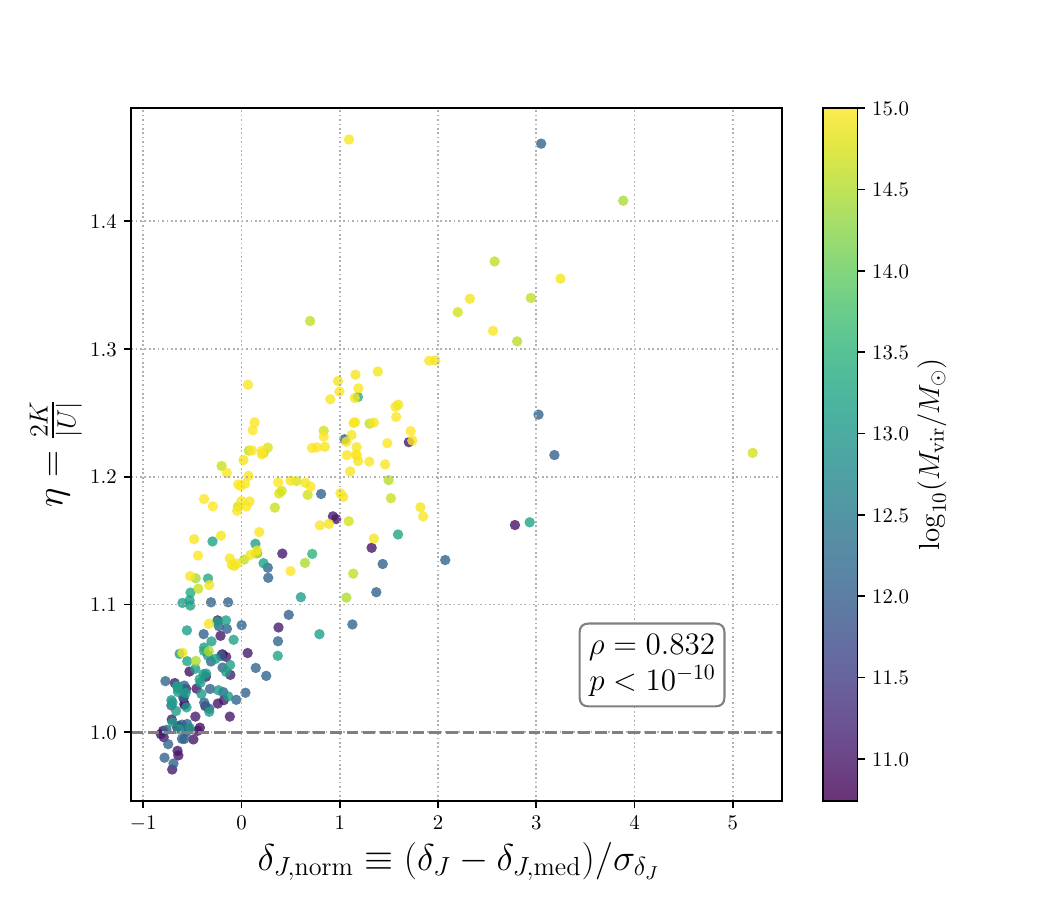}
        \caption{The relation between virial ratio and total normalized Jeans deviation parameter for host haloes across all \textit{Symphony} suites. The horizontal dashed line marks virial equilibrium and the legend lists the Spearman correlation coefficients.}
        \label{fig:jenas_vs_eta}
    \end{minipage}
    \hfill 
    \begin{minipage}{0.48\textwidth}
        \centering
        \renewcommand{\thefigure}{C1}
        \includegraphics[width=\linewidth]{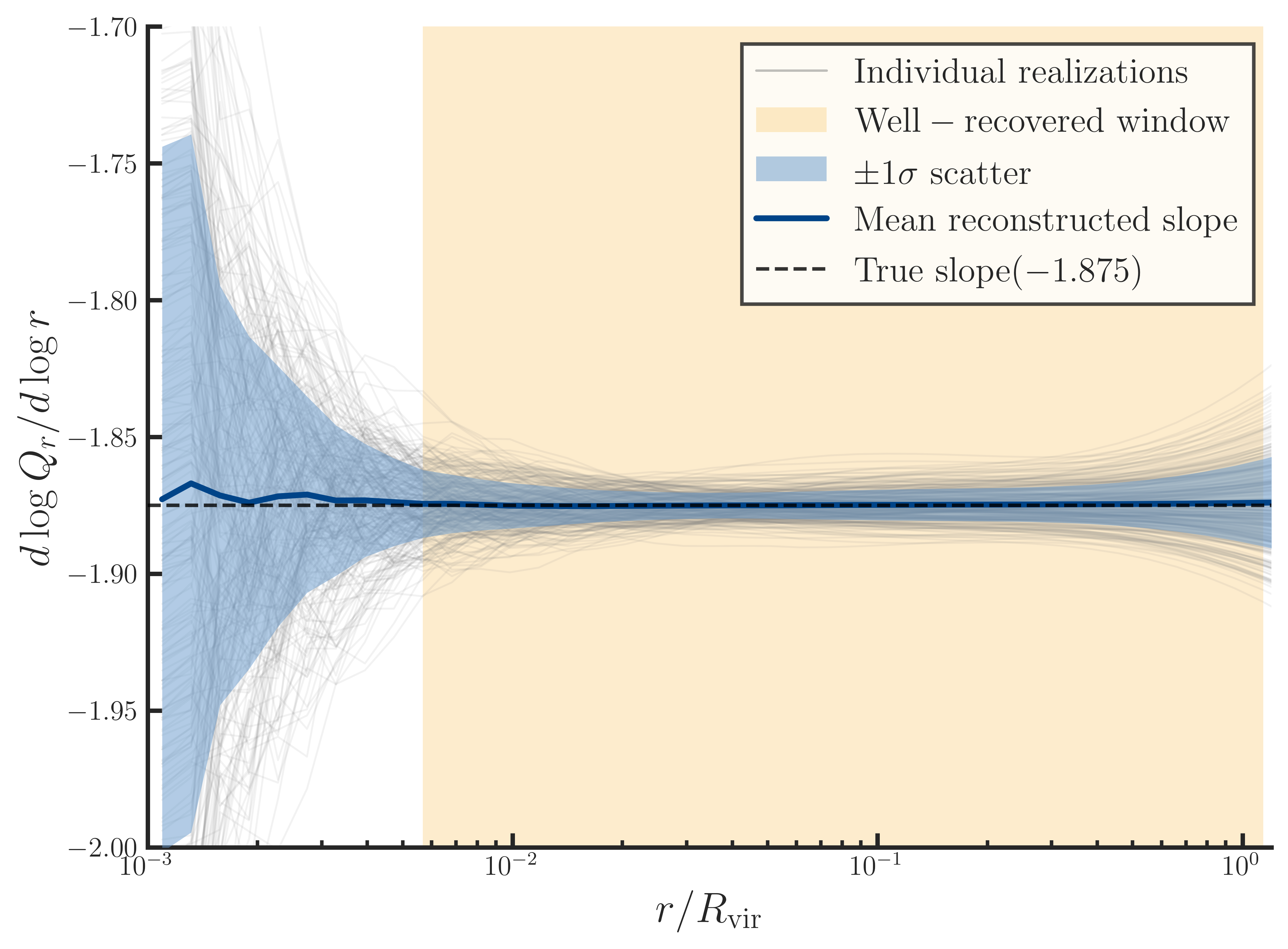}
        \caption{Recovery test of the PPSD slope using the \texttt{constant\_jerk} Kalman smoothing method. The thin solid lines show the 250 Poisson-noised realizations of mock NFW haloes, and the thick solid line presents the mean recovered slope (solid line) with $\pm1\sigma$ scatter (shaded).}
        \label{fig:recovery_test}
    \end{minipage}
\end{figure*}

The dynamical equilibrium of dark matter haloes can be characterized using several criteria, such as the substructure mass fraction, the offset between the center of mass and the potential minimum, and the virial ratio (e.g., \citealt{Neto:2007vq}). In this work, we introduce the Jeans deviation parameter, defined in Equation~\ref{eq:jeans_dev}, which provides a more direct measure of the degree to which a system satisfies the Jeans equilibrium condition. This quantity can be evaluated as a function of radius to quantify the local imbalance between pressure support and gravitational force.

To verify that the total Jeans deviation parameter captures a similar physical meaning as conventional equilibrium measures, we compare it to the commonly used virial ratio for each host halo in the \textit{Symphony} simulations (Figure~\ref{fig:jenas_vs_eta}). The two quantities exhibit a strong and statistically significant correlation (Spearman coefficient $\rho = 0.820$, $p < 10^{-10}$), confirming that the total Jeans deviation parameter provides a reliable estimator of a halo’s dynamical relaxation state. For reference, the majority of \textit{Symphony} host haloes satisfy the virial ratio criterion for relaxed haloes adopted by \citet{Neto:2007vq}, i.e., $\eta < 1.35$.

\section{Smoothing derivative algorithm}
\label{sec:smooth_derivative}

To assess the reliability of our slope measurement method, we perform a recovery test using mock PPSD profiles. Specifically, we construct $Q_r$ profiles that follow an exact power-law relation $Q \propto r^{-1.875}$ and introduce statistical noise by Poisson-sampling the expected particle counts within each radial bin, matching the noise level of our simulations.

Figure~\ref{fig:recovery_test} presents the mean recovered PPSD slope and its $\pm1\sigma$ scatter, measured using 250 individual noisy realizations (close to the number of hosts in \textit{Symphony}) for an NFW halo with concentration $c=10$. For each realization, the slope is obtained using the \texttt{constant\_jerk} variant of the Kalman smoothing algorithm. The yellow-shaded radial range indicates the ``well-recovered window,'' defined as the radial range where the mean recovered slope deviates by less than $1\%$ from $-1.875$ and the realization-to-realization scatter remains below the mean dispersion across all radii, corresponding to $6\times10^{-3}<r/R_{\text{vir}}<1.1$. Throughout this work, we therefore restrict our slope fitting to this range to ensure noise-robust measurements.

\section{Suite consistency tests}
\label{sec:suites_consistency}

\begin{figure*}
    \centering
    \includegraphics[width=1\linewidth]{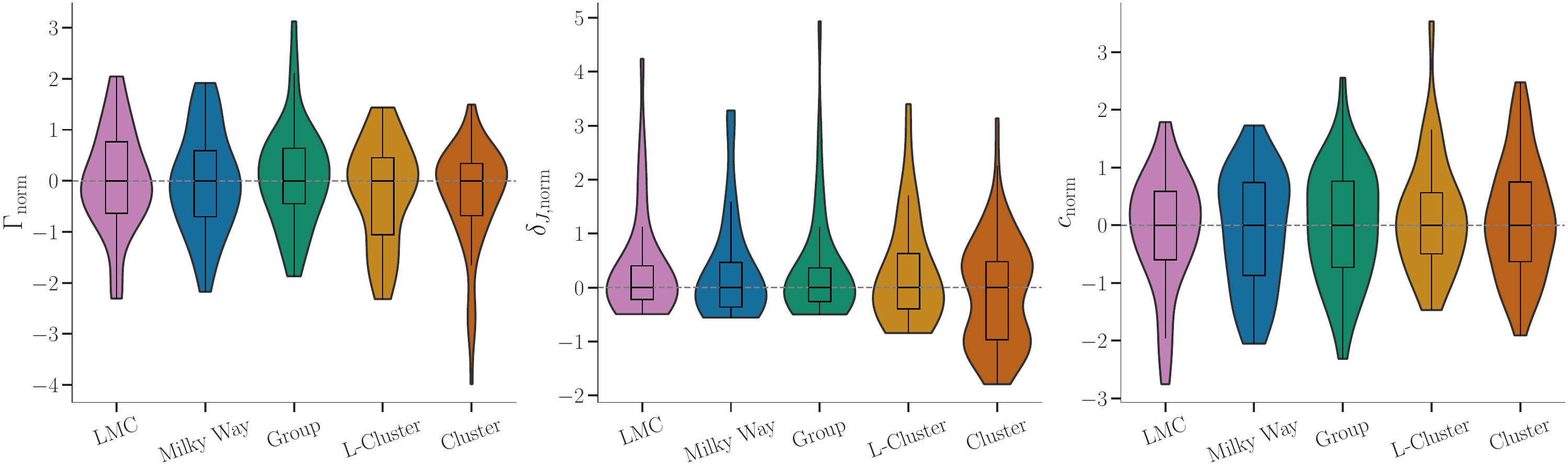}
    \caption{
    Violin and box plots of the normalized concentration, accretion rate, and total Jeans deviation parameter for host haloes in each \textit{Symphony} suite. 
    Each distribution is normalized within its own suite following the manner of Equation~\ref{eq:normalize}, with violin shapes showing the full range and boxes indicating the interquartile ranges and medians.
    }
    \label{fig:suites_consistency}
\end{figure*}

Halo properties such as concentration, accretion rate, and the total Jeans deviation parameter are expected to depend weakly on halo mass (e.g., \citealt{Mandelbaum_2008}, \citealt{McBride_2009}, \citealt{Ludlow_2014}). To isolate correlations with the PPSD slope, we have normalized these secondary halo properties in certain analyses using Equation~\ref{eq:normalize}. To verify this procedure,
Figure~\ref{fig:suites_consistency} presents the distributions of the normalized concentration, accretion rate, and total Jeans deviation parameter for host haloes in each \textit{Symphony} suite. The normalized concentration ($c_{\rm norm}$) and accretion rate ($\Gamma_{\rm norm}$) show nearly identical distributions across all suites. Kolmogorov–Smirnov (KS) tests further confirm this consistency, with $p$-values greater than 0.05 for all pairwise suite comparisons. 

In contrast, the distribution of the normalized Jeans deviation parameter ($\delta_{J,\rm norm}$) in the Cluster suite differs significantly from the others, as indicated by the KS test and by its bimodal structure in Figure~\ref{fig:suites_consistency}. This is not surprising, given the major recent accretion events that many hosts in our Cluster sample experience~\citep{Nadler_2023}; this analysis justifies our choice in Figure~\ref{fig:slope_vs_jeans}, to only include the LMC, Milky Way, Group, and L-Cluster suites.


\bsp	
\label{lastpage}
\end{document}